\documentclass[aps,preprint,eqsecnum]{revtex4}
\usepackage{epsfig}
\usepackage{graphicx}
\usepackage{psfrag}
\newcommand{\be}{\begin{eqnarray}}
\newcommand{\ee}{\end{eqnarray}}

\def\nn{\nonumber}

\def\la{\lambda}

\newcommand\as{\alpha_s}
\newcommand\f[2]{\frac{#1}{#2}}

\def\d{{\rm d}}

\def\xa{x_{1}}
\def\xb{x_{2}}
\def\xaa{x_{1}^{0}}
\def\xbb{x_{2}^{0}}

\begin{document}

\preprint{BNL-NT-06/3}
\preprint{RBRC-580}
\preprint{hep-ph/0601162}
\vspace*{1cm}
\title
{Threshold Resummation for $W$-Boson Production at RHIC\\[1cm] } 
\author{\bf A. Mukherjee}
\email{asmita@phy.iitb.ac.in}
\affiliation{Physics Department, Indian Institute of Technology Bombay,\\
Powai, Mumbai 400076, India.}
\author{\bf W. Vogelsang}\email{vogelsan@quark.phy.bnl.gov }
\affiliation{Physics Department and RIKEN-BNL Research Center,
Brookhaven National Laboratory,\\
Upton, New York 11973, U.S.A.}

\date{\today\\[2cm]}

\begin{abstract}
We study the resummation of large logarithmic perturbative corrections
to the partonic cross sections relevant for the process 
$pp\to W^{\pm} X$ at the BNL Relativistic Heavy Ion Collider (RHIC).
At RHIC, polarized protons are available, and spin asymmetries for 
this process will be used for precise measurements of the 
up and down quark and anti-quark distributions in the proton.  
The corrections arise near the threshold for the partonic 
reaction and are associated with soft-gluon emission. We perform
the resummation to next-to-leading logarithmic accuracy,
for the rapidity-differential cross section. We find that
resummation leads to relatively moderate effects on the 
cross sections and spin asymmetries. 
\end{abstract}
\maketitle

\section{Introduction}
The exploration of the spin structure of the nucleon has
by now become a classic topic of nuclear and particle physics.
Most of our knowledge so far comes from spin-dependent 
deeply-inelastic scattering (DIS) which has produced spectacular
results~\cite{reviews}, the single most important one being that 
the total quark spin contribution to the nucleon spin is only 
about $20-25\%$. This has led to a new generation of experiments
that aim at further unraveling the spin structure of the nucleon.
Among these are experiments at the Relativistic Heavy Ion Collider (RHIC)
at BNL. Here, a new method is used, namely to study collisions of 
two polarized protons~\cite{rhicrev}. This allows in particular 
clean probes of the polarized
gluon distribution in the proton, whereby one is hoping to 
learn about the contribution of gluons to the proton spin. 

DIS has given us exciting information on the spin-dependent
quark and anti-quark distributions $\Delta q(x,\mu^2)$, $\Delta 
\bar{q}(x,\mu^2)$, where
$\Delta q(x,\mu^2)\equiv q^{\uparrow}(x,\mu^2)-q^{\downarrow}(x,\mu^2)$
with $q^{\uparrow}(x,\mu^2)$ ($q^{\downarrow}(x,\mu^2)$) denoting the 
distributions of quarks with positive (negative) helicity and light-cone
momentum fraction $x$ in a proton with positive helicity, at 
factorization scale $\mu$. Inclusive DIS via photon exchange, however, 
gives access only to the combinations $\Delta q+\Delta 
\bar{q}$. In order to understand the proton helicity structure in detail, 
one would like to learn about the various quark and anti-quark densities, 
$\Delta u, \Delta \bar{u}, \Delta d, \Delta \bar{d},\Delta s, 
\Delta \bar{s}$, {\it individually}. This is for example relevant for
comparisons to models of nucleon structure which generally make 
clear qualitative predictions about, for example, the flavor 
asymmetry $\Delta \bar{u}-\Delta \bar{d}$ in the proton 
sea~\cite{su2}. These predictions are often related to 
fundamental concepts such as the Pauli principle: since
the proton has two valence-$u$ quarks which primarily spin 
along with the proton spin direction, $u\bar{u}$ pairs
in the sea will tend to have the $u$ quark polarized opposite to 
the proton. Hence, if such pairs are in a spin singlet, one 
expects $\Delta \bar{u}>0$ and, by the same reasoning, $\Delta 
\bar{d}<0$. Such questions become all the more exciting 
due to the fact that rather large {\it unpolarized}
asymmetries $\bar{u}-\bar{d}\neq 0$ have been observed in DIS and 
Drell-Yan measurements~\cite{dysu2,dysu21,dysu22}. 

There are two known ways to distinguish between 
quark and anti-quark polarizations experimentally, and also to 
achieve at least a partial flavor separation. Semi-inclusive 
measurements in DIS are one possibility, explored by 
SMC \cite{smc} and, more recently
and with higher precision, by {\sc Hermes} \cite{hermesdq}.
Results are becoming available now also from the 
{\sc Compass} experiment~\cite{compass}, and measurements
have been proposed for the Jefferson Laboratory~\cite{jlab}.
One detects a hadron $h$ in the final state, so that instead of $\Delta q+
\Delta \bar{q}$ the polarized DIS cross section becomes 
sensitive to $\;\Delta q(x)\,D_q^h(z) + \Delta  \bar{q}(x)\,
D_{\bar{q}}^h (z)$, for a given quark flavor. Here, the $D_i^h(z)$ are 
fragmentation functions, with $z$ denoting the fraction of the momentum of
the struck quark transfered to the hadron. The dependence on the details 
of the fragmentation process limits the accuracy of this method. 
There is much theory activity currently on SIDIS, focusing on
next-to-leading order (NLO) corrections, target fragmentation, and higher twist
contributions \cite{sidis}. 

The other method to learn about the $\Delta q$ and $\Delta \bar{q}$
will be used at RHIC. Here one considers spin asymmetries
in the production of $W^{\pm}$ bosons~\cite{BOURRELY93}. 
Because of the 
violation of parity in the coupling of $W$ bosons to quarks and 
anti-quarks, {\it single}-longitudinal spin asymmetries, defined as 
\begin{equation}
A_{L}\equiv  \frac{\d \sigma^{++}
+ \d \sigma^{+-} - \d \sigma^{-+} - \d \sigma^{--}}{\d \sigma^{++}
+ \d \sigma^{+-} + \d \sigma^{-+} + \d \sigma^{--}}\equiv
\frac{\d \Delta \sigma}{\d \sigma} \; ,
\label{eq:defl}
\end{equation}
can be non-vanishing. Here the $\sigma$ denote cross sections
for scattering of protons with definite helicities as indicated
by the superscripts; as one can see, we are summing over
the helicities of the second proton, which leads to the single-spin
process $\vec{p}p\to W^{\pm}X$. Since the $W$ mass sets a large
momentum scale in the process, the cross sections factorize
into convolutions of parton distribution functions and partonic
hard-scattering cross sections. The latter are amenable to QCD
perturbation theory, the lowest-order reaction being the ``Drell-Yan''
process $q\bar{q}'\to W^{\pm}$. For produced $W^+$, the dominant 
contributions come from $u\bar{d}\to W^+$, and the spin asymmetry is 
approximately given by~\cite{rhicrev,craigie83,BOURRELY93}
\begin{equation}
A_{L}^{W^+} 
= \frac{\Delta u(x_1^0,M_W^2) \bar{d}(x_2^0,M_W^2) - 
\Delta \bar{d} (x_1^0,M_W^2) u (x_2^0,M_W^2) } 
{u(x_1^0,M_W^2) \bar{d}(x_2^0,M_W^2) + 
\bar{d} (x_1^0,M_W^2) u (x_2^0,M_W^2) } \; . \label{asyw+}
\end{equation} 
Here, as indicated, the parton distributions will be probed at a scale
$\sim M_W$, with $M_W$ the $W$ mass, 
and the momentum fractions $x_1^0$ and $x_2^0$ are related 
to the rapidity $\eta$ of the $W^+$ by 
\begin{equation}
x_{1,2}^0 = \frac{M_W}{\sqrt{S}}\, e^{\pm \eta}\; ,
\label{eq:x0def}
\end{equation}
where $\sqrt{S}$ is the $pp$ center-of-mass 
energy which at RHIC is 200 or 500~GeV. Note that $\eta$ is counted
positive in the forward region of the polarized proton. From 
Eq.~(\ref{asyw+}) it follows that at large $\eta$, where 
$x_1^0 \sim 1$ and $x_2^0\ll 1$, the asymmetry will
be dominated by the valence distribution probed at $x_1^0$, 
and hence give direct access
to $\Delta u(x_1^0,M_W^2)/u(x_1^0,M_W^2)$. Likewise, for large negative
$\eta$, $A_L^{W^+}$ 
is given by $-\Delta \bar{d}(x_1^0,M_W^2)/\bar{d}(x_1^0,M_W^2)$. 
For negatively charged $W$ bosons, one finds correspondingly 
sensitivity to $\Delta d(x_1^0,M_W^2)/d(x_1^0,M_W^2)$ and 
$-\Delta \bar{u}(x_1^0,M_W^2)/\bar{u}(x_1^0,M_W^2)$ at large positive
and negative $W$ rapidities, respectively. This is the key idea
behind the planned measurements of 
$\Delta u, \Delta \bar{u}, \Delta d, \Delta \bar{d}$ at RHIC.

In practice, a significantly more involved strategy needs to be
used in order to really relate the single-longitudinal spin
asymmetries to the polarized quark and anti-quark densities. 
Partly, this is an experimental issue: the detectors at RHIC
are not hermetic, which means that missing-momentum techniques
for the charged-lepton neutrino ($l\nu_l$) 
final states cannot be straightforwardly used to detect the $W$
and reconstruct its rapidity. There are, however, workarounds
to this problem. One is to assume that the $W$ is dominantly
produced with near-zero transverse momentum, in which case one
can relate the measured rapidity of the charged lepton to
that of the $W$, up to an irreducible sign ambiguity~\cite{bland,NadYuan2}.
Ultimately, it may then become more expedient not even to consider
the $W$ rapidity, but to formulate all observables directly in terms 
of the charged-lepton rapidity. This approach has been studied in 
detail in~\cite{NadYuan2,NadYuan1}. It was found that despite the fact
that the interpretation of the spin asymmetries in terms of
the $\Delta u, \Delta \bar{u}, \Delta d, \Delta \bar{d}$
becomes more involved, there still is excellent sensitivity
to them. Contributions by $Z$ bosons need to be taken into account 
as well. Very recently, it has also been proposed~\cite{NadTalk} to 
use hadronic $W$ decays in the measurements of $A_L$. These
have the advantage that one just needs to look for events with two
hadronic jets. A potential drawback is that, while the QCD 
two-jet background cancels in the parity-violating numerator of
$A_L$, it does contribute to the denominator and therefore will likely
reduce the size of the spin asymmetries. 

There are also theoretical issues that modify the picture 
related to Eq.~(\ref{asyw+}) described above. There are for example
Cabibbo-suppressed contributions. It is relatively straightforward to
take these into account, even though one needs to keep in mind that
they involve the polarized and unpolarized strange quark distributions.
More importantly, there are higher-order QCD corrections to the 
leading order (LO) process $q\bar{q}'\to W^{\pm}$. At next-to-leading order, 
one has the partonic reactions $q\bar{q}'\to W^{\pm}g$ and 
$qg\to W^{\pm}q'$. For the unpolarized and the single-longitudinal 
spin cases, the cross sections for these have been given and studied 
in Refs.~\cite{aem,kubar} 
and~\cite{ratcliffe,weber,weberqt,kamal,tgdy,gehrmann,grh}, 
respectively. In the present paper, we will improve the theoretical framework
by going beyond NLO and performing a resummation of certain
logarithmically enhanced terms in the partonic cross sections
to all orders in perturbation theory.

The corrections we are considering arise near ``partonic threshold'', 
when the initial partons have just enough energy to produce the
$W$ boson. Here the phase space available for real-gluon radiation 
vanishes, while virtual corrections are fully allowed. The cancellation
of infrared singularities between the real and virtual diagrams then
results in large logarithmic ``Sudakov'' corrections to 
the LO $q\bar{q}'$ cross section. For example, for the cross section 
integrated over all rapidities of the $W$, the most important logarithms
(the ``leading logarithms'' (LL)) take the form $\as^k [ 
\ln^{2k-1}(1-z)/(1-z)]_+$ at the $k$th order of perturbation
theory, where $z=M_W^2/\hat{s}$ with
$\hat{s}$ the partonic center-of-mass energy, and
$\as$ is the strong coupling. The ``$+$''-distribution 
is defined in the usual way and regularizes the infrared 
behavior $z\to 1$. Subleading logarithms are down by one or 
more powers of the logarithm and are referred to 
as ``next-to-leading logarithms'' (NLL), and so 
forth. Because of the interplay of the partonic cross sections 
with the steeply falling parton distributions, the threshold
regime can make a substantial contribution to the cross section
even if the hadronic process is relatively far from threshold,
that is, even if $M_W^2/S\ll 1$. If $M_W^2/S\sim 1$, the threshold
region will completely dominate the cross section. This is 
expected to be the case for $W$ production at 
the RHIC energies, in particular
at $\sqrt{S}=200$~GeV. Sufficiently close to partonic threshold, 
the perturbative series will be only useful if the large logarithmic 
terms are taken into account to all orders in $\as$. This is achieved 
by threshold resummation. 

Drell-Yan type processes have been the first for which threshold 
resummation was derived. The seminal work in~\cite{dyresum,dyresum2}
has been the starting point for the developments of related 
resummations for many hard processes in QCD~\cite{f2}. The resummation for
Drell-Yan is now completely known to next-to-next-to-leading 
logarithmic (NNLL) accuracy~\cite{f1}.  
In this paper we will, however, perform the resummation only
at NLL level. Since the strong coupling at scale $M_W$ is 
relatively small, we expect this to be completely sufficient for
a good theoretical description. In addition, away from the threshold
region, one needs to match the resummed calculation to the available 
fixed-order one. In our case of single-spin production of $W$'s, this
is NLO. A consistent matching of a NNLL resummation would require 
matching to NNLO, which is not yet available. 

As we discussed above, of particular interest at RHIC for determining
the spin-dependent quark and anti-quark densities are the rapidity 
distributions of the $W$ bosons. In order to present phenomenology 
relevant for RHIC, we will therefore perform the resummation
for the rapidity-dependent cross sections. Note that we will
concentrate only on the case of the distributions in $W$-boson
rapidity; from the earlier discussion it follows that for future 
studies it would be even more desirable to consider the distributions
in charged-lepton rapidities. To treat the resummation at fixed
$W$ rapidity, we will employ the technique developed in~\cite{sv}. 
This entails to use Mellin moments in $\tau=M_W^2/S$ of the cross 
section, as is customary in threshold resummation, but also 
a Fourier transform in rapidity. In~\cite{sv}, this method was 
applied to the prompt-photon cross section. For the case of the 
Drell-Yan cross section it simplifies considerably. We note that 
techniques for the threshold resummation of the Drell-Yan cross 
section at fixed rapidity were also briefly discussed 
in~\cite{LaenenSterman}. We verify the main result of that paper, 
and for the first time, present phenomenological results for this case. 

Before turning to the main part of the paper, we mention that 
important work on another type of resummation of the $W$ production 
cross sections at RHIC has been performed in the literature, namely
for the transverse-momentum distribution ($p_T$) of the 
$W$'s~\cite{weberqt,NadYuan2,NadYuan1}. 
At lowest order, the $W$ is produced with vanishing transverse momentum. 
Gluon radiation generates a recoil transverse momentum. By a similar 
reasoning as above, when $p_T$ tends to zero, large logarithmic
corrections develop in the $p_T$ spectrum of the $W$'s. These
can be resummed as well, which was studied for the case of RHIC in great 
detail in~\cite{weberqt,NadYuan2,NadYuan1}. For 
the total or the rapidity-differential 
$W$ cross section we are interested in, $p_T$ is integrated, and 
the associated large logarithms turn partly into threshold
logarithms and partly into nonlogarithmic terms. Therefore, the 
threshold logarithms we are considering here become the main
source of large corrections to the cross section. We note that
ultimately it would be desirable to perform a ``joint'' resummation
of the $p_T$ and threshold logarithms, as developed in Ref.~\cite{LSV}.

The remainder of this paper is organized as follows. In Section~\ref{sec2}
we present general formulas for the single-spin cross section for $W$ 
production as a function of rapidity, and also discuss the NLO
corrections. In Section~\ref{sec3}, we give details of the
Mellin and Fourier transforms that are 
useful in achieving threshold resummation of
the rapidity-dependent cross section. The next two sections provide 
the formulas for the resummed cross sections. In Section~\ref{sec6} 
we present our numerical results for RHIC.

\section{Cross section for $W$ production in $\vec{p}p$ collisions \label{sec2}}

The rapidity-differential Drell-Yan cross section for $W^{\pm}$ production in 
singly-polarized $pp$ collisions can be written as \cite{gehrmann}
\begin{eqnarray}
\frac{\d \Delta \sigma}{\d \eta} & = & {\cal N} \sum_{i,j} 
\int_{x_1^0}^1 \d x_1 \int_{x_2^0}^1 \d x_2 \,
\Delta {\cal D}_{ij}\left(x_1,x_2,\xaa,\xbb,\alpha_s(\mu^2),
\frac{M_W^2}{\mu^2}\right)\, \Delta f_i(x_1,\mu^2)\,
f_j(x_2,\mu^2) \nonumber \\
& = & {\cal N} \sum_{i,j} c_{ij}
\int_{x_1^0}^1 \d x_1 \int_{x_2^0}^1 \d x_2 \nonumber \\
& &  \hspace{-1.6cm}
\times \Bigg\{
D_{q\bar{q}}
\left(x_1,x_2,\xaa,\xbb,\alpha_s(\mu^2),
\frac{M_W^2}{\mu^2}\right)
\Big[ - \Delta q_i(x_1,\mu^2) \bar{q}_j(x_2,\mu^2) +
\Delta \bar{q}_i(x_1,\mu^2) q_j(x_2,\mu^2) \Big] \nonumber \\
& & \hspace{-1.2cm} + 
\Delta D_{gq} \left(x_1,x_2,\xaa,\xbb,\alpha_s(\mu^2),
\frac{M_W^2}{\mu^2}\right) \Delta G(x_1,\mu^2) \left[
q_j(x_2,\mu^2) -
\bar{q}_j (x_2,\mu^2) \right] \nonumber \\
& & \hspace{-1.2cm} +  
D_{qg} \left(x_1,x_2,\xaa,\xbb,\alpha_s(\mu^2),
\frac{M_W^2}{\mu^2}\right) \left[
- \Delta q_i(x_1,\mu^2) +
\Delta \bar{q}_i (x_1,\mu^2) \right]
G(x_2,\mu^2)  \Bigg\}\; , \label{eq:lmas}
\end{eqnarray}
where the $\Delta f_i$, $f_j$ are the polarized and unpolarized
parton distributions, respectively, and where in the second equation
we have written out explicitly the various contributions that are possible 
through NLO. Furthermore, the normalization factor ${\cal N}$ is given by
\begin{equation}
{\cal N} = \frac{\pi G_F M_W^2\sqrt{2}}{3 S}\; ,
\end{equation}
with the Fermi constant $G_F$, and $c_{ij}$ are the coupling 
factors for $W^{\pm}$ bosons, 
\begin{eqnarray}
& c_{ij} = |V_{ij}|^2
\end{eqnarray}
with $V_{ij}$ the CKM mixing factors for the quark flavors $i,j$.
$\mu\sim M_W$ denotes the renormalization/factorization scales which we 
take to be equal. $x_1^0$ and $x_2^0$ have been defined above 
in Eq.~(\ref{eq:x0def}).

The $D_{q\bar{q}}$, $\Delta D_{gq}$, $D_{qg}$ in Eq.~(\ref{eq:lmas})
are the hard-scattering functions. For the single-spin cross section,
always one initial parton is unpolarized. Therefore, thanks to helicity
conservation in QCD and the $V-A$ structure of the $Wq\bar{q}'$ vertex, 
when the polarized parton is a quark, one finds that the spin-dependent
partonic cross section is the negative of the unpolarized one. We
have therefore omitted the $\Delta$'s in these cases and only
included a $\Delta$ for the case of an initial polarized gluon
where there is no trivial relation between the polarized and the 
unpolarized cross sections. Each of the hard-scattering functions $D_{ij}$
(or $\Delta D_{ij}$)
is a perturbative series in the strong coupling $\alpha_s(\mu^2)$:
\begin{eqnarray}
D_{ij}\left(x_1,x_2,\xaa,\xbb,\alpha_s(\mu^2),
\frac{M_W^2}{\mu^2}\right)&=&D_{ij}^{(0)}\left(x_1,x_2,\xaa,\xbb\right)
\nonumber\\ 
&&+\frac{\alpha_s(\mu^2)}{2\pi}D_{ij}^{(1)}\left(x_1,x_2,\xaa,\xbb,
\frac{M_W^2}{\mu^2}\right)+\ldots \; .
\end{eqnarray}
Only the $q\bar{q}'$ process has a lowest-order [${\cal O}(\alpha_s^0)$]
contribution:  
\begin{eqnarray}
D_{q\bar q}^{(0)} (x_1,x_2,\xaa,\xbb) & = & \delta (x_1-x_1^0)\,
\delta (x_2-x_2^0) \; , \nonumber \\
\Delta D_{gq}^{(0)}(x_1,x_2,\xaa,\xbb) &=& D_{qg}(x_1,x_2,
\xaa,\xbb)\;=\; 0 \; .
\label{locoeff}
\end{eqnarray}
The NLO hard-scattering functions have rather lengthy expressions which have
been derived in Refs.~\cite{kubar,weber,tgdy}. For the reader's
convenience we collect them in the Appendix. As can be seen from 
Eqs.~(\ref{locoeff}), 
(\ref{app1})-(\ref{app5}), the coefficients contain distributions in 
$(x_1-x_1^0)$ and $(x_2-x_2^0)$. These are the terms addressed by 
threshold resummation. More precisely, it turns out that only 
products of two ``plus''-distributions, or a product of 
a ``plus''-distribution and a delta-function, are leading 
near threshold. This occurs only for the coefficient $D_{q\bar{q}}$. 
To be able to write down the resummed expressions for arbitrary rapidity, 
we need to take suitable integral transforms, to which we shall turn now.  

\section{Mellin and Fourier transforms, and threshold limit \label{sec3}}

Consider the production of a $W$ boson through a single generic
partonic reaction involving initial partons $i$ and $j$. We
define a double transform of the cross section, in terms of a Mellin transform
in $\tau=M_W^2/S$ and a Fourier transform in the rapidity of the $W$ boson:
\be
\Delta 
\tilde{\sigma}(N,M)&\equiv& \int_0^1 d \tau \, \tau^{N-1} 
\int_{-\ln\frac{1}{\sqrt{\tau}}}^
{\ln\frac{1}{\sqrt{\tau}}} d \eta\, 
{\mathrm{e}}^{i M \eta} \,
{d\Delta \sigma \over d \eta}  \\
&&\hspace*{-12mm}=\, {\cal N}\int_0^1 dx_1 x_1^{N+i M/2} \Delta 
f_i (x_1) \int_0^1 dx_2 \,x_2^{N-iM/2} 
f_j (x_2)\int_0^1 dz \,z^{N-1} \int_{-\ln\frac{1}{\sqrt{z}}}^
{\ln\frac{1}{\sqrt{z}}} d {\hat \eta} \, 
{\mathrm{e}}^{i M {\hat \eta}}\, \Delta {\cal D}_{ij} \; . \nonumber
\label{cross1}
\ee
Here we have suppressed the argument of the partonic cross section
as well as the scale dependence of the parton distributions.
We have introduced  parton level variables 
$z=\tau/x_1 x_2$ and ${\hat \eta}=\eta -{1\over 2} \ln(x_1/x_2)$.
Each of the functions $\Delta {\cal D}_{ij}$ is of the form
$(x_1 x_2)^{-1}$ times a function of $x_1^0/x_1$ and $x_2^0/x_2$
only. Defining the moments 
\begin{equation}
f^N (\mu^2) \equiv \int_0^1 dx \,x^{N-1} f(x, \mu^2) \; , \nonumber\\
\end{equation}
of the parton densities, and 
\begin{equation}
\Delta \tilde{{\cal D}}_{ij}(N,M)\equiv
\int_0^1 dz \,z^{N-1} \int_{-\ln\frac{1}{\sqrt{z}}}^
{\ln\frac{1}{\sqrt{z}}} d {\hat \eta} \, 
{\mathrm{e}}^{i M {\hat \eta}}\, \left( x_1 x_2 \Delta {\cal D}_{ij} 
\right)\; ,
\label{Ddef}
\end{equation}
we therefore have:
\begin{equation}
\Delta 
\tilde{\sigma}(N,M)= {\cal N}\,\Delta f_i^{N+i M/2}\,f_j^{N-iM/2}
\,\Delta \tilde{{\cal D}}_{ij}(N,M)\; .  
\label{cross2}
\end{equation}

At lowest order,
\begin{equation}
x_1 x_2 D_{q\bar{q}}^{(0)}\left(x_1,x_2,\xaa,\xbb\right) =x_1 x_2
\delta (x_1-x_1^0)\,\delta (x_2-x_2^0)=
\delta \left(1-\sqrt{z}{\mathrm{e}}^{\hat{\eta}}
\right)\,\delta\left(1-\sqrt{z}{\mathrm{e}}^{-\hat{\eta}}\right)\; ,
\end{equation}
and we find
\begin{equation}
\int_0^1 dz \,z^{N-1} \int_{-\ln\frac{1}{\sqrt{z}}}^
{\ln\frac{1}{\sqrt{z}}} d {\hat \eta} \, 
{\mathrm{e}}^{i M {\hat \eta}}\,\delta 
\left(1-\sqrt{z}{\mathrm{e}}^{\hat{\eta}}
\right)\delta\left(1-\sqrt{z}{\mathrm{e}}^{-\hat{\eta}}\right)\,=\,
\int_0^1 dz \,z^{N-1}\,2\cos\left( M\ln\frac{1}{\sqrt{z}}\right)\,
\delta(1-z)\; .\label{lomom}
\end{equation}
As one can see, a $\delta(1-z)$ emerges, as expected from 
the well-known expression for the rapidity-integrated LO Drell-Yan
cross section~\cite{aem}. 
Beyond LO, a closed calculation of the double moments 
$\Delta \tilde{{\cal D}}_{ij} (N,M)$ is very difficult. For example, 
at NLO, one would need to use the expressions given in 
Eqs.~(\ref{app1})-(\ref{app5})
and take the moments in $N$ and $M$. Fortunately, a great
simplification occurs in the near-threshold limit. 
Equation~(\ref{lomom}) shows that the partonic threshold,
after taking Fourier moments in rapidity, is reached at $z=1$.
In Mellin-moment space, this corresponds to the large-$N$
limit. Even though the Cosine factor is obviously unity 
in conjunction with the term $\delta(1-z)$, it is generic, as we will
now see.

At NLO, one finds from Eqs.~(\ref{app1})-(\ref{app5})
the following structure near threshold:
\begin{eqnarray}
&&\hspace*{-2cm}
\int_0^1 dz \,z^{N-1} \int_{-\ln\frac{1}{\sqrt{z}}}^
{\ln\frac{1}{\sqrt{z}}} d {\hat \eta} \, 
{\mathrm{e}}^{i M {\hat \eta}}\, x_1 x_2 D_{q\bar{q}}^{(1)}
\left(x_1,x_2,\xaa,\xbb,\alpha_s(\mu^2),
\frac{M_W^2}{\mu^2}\right) \nonumber \\
&=&\int_0^1 dz \,z^{N-1}\,\Bigg\{\,2\cos\left( M\ln\frac{1}{\sqrt{z}}\right)\,
C_F\left[ 8 \left( \frac{\ln (1-z)}{1-z}\right)_+ +
\frac{4}{1-z}_+ \ln
\frac{M_W^2}{\mu^2}  \right.\nonumber \\
&&\left. +\left( -8+\frac{2 \pi^2}{3}+
3\ln \frac{M_W^2}{\mu^2}\right) \delta(1-z) \right] +
{\cal O}(1-z) \Bigg\}\; ,
\label{zterms}
\end{eqnarray}
where $C_F=4/3$. The factor in square brackets is the large-$z$ limit of
the well-known~\cite{aem} ${\cal O}(\alpha_s)$ QCD correction 
to the rapidity-integrated Drell-Yan cross section through the 
$q\bar{q}$ channel. As usual, the ``plus''-distributions are 
defined over the integral from 0 to 1 by
\begin{equation}
\int_0^1 dz f(z) \left[ g(z)\right]_+\equiv
\int_0^1 dz \left(f(z)-f(1)\right) g(z) \; .
\end{equation} 
As before, the Cosine factor emerges, which is 
subleading near threshold since 
\begin{equation}
\cos\left( M\ln\frac{1}{\sqrt{z}}\right)=
1 - \frac{(1-z)^2 M^2}{8} + {\cal O}\left((1-z)^4 M^4 \right)\; .
\label{costerms}
\end{equation}
It is therefore expected to be a good approximation to set this term
to unity,
and in any case consistent with the threshold approximation. 
On the other hand, the term carries information on rapidity through
the Fourier variable $M$. At very large $M$ keeping the Cosine term
may be more important. In the following, we will ignore the term
and just point out what changes to our formulas below would occur
if one kept it. We will also study the numerical relevance of the
Cosine term later in the phenomenology section. 

We now take the Mellin moments of the expression in Eq.~(\ref{zterms}).
At large $N$, in the near-threshold limit, the moments of the NLO correction become
\begin{eqnarray} \label{c1qlargen}
&&\hspace*{-1cm}\int_0^1 dz \,z^{N-1}\,
C_F\left[ 8 \left( \frac{\ln (1-z)}{1-z}\right)_+ +
\frac{4}{1-z}_+ \ln
\frac{M_W^2}{\mu^2} + \left( -8+\frac{2 \pi^2}{3}+
3\ln \frac{M_W^2}{\mu^2}\right) \delta(1-z) \right] \nonumber \\
&=& 2 C_F \left[
2 \ln^2 (\bar{N})  + \left( \frac{3}{2} - 2 \ln(\bar{N}) \right)\ln  
\frac{M_W^2}{\mu^2}- 4 + \frac{2\pi^2}{3}  \right] +
{\cal O}\left( \frac{1}{N}\right)\; ,
\end{eqnarray}
where
\be
\bar{N}=N {\rm e}^{\gamma_E} \; .
\ee
The main result, expressed by Eq.~(\ref{c1qlargen}), 
is that near threshold the double moments of the 
rapidity-dependent cross section are independent 
of the Fourier (conjugate to rapidity) variable $M$,
up to small corrections that are suppressed near threshold. 
The $N$-dependence is identical to that of the rapidity-integrated 
cross section. This result was also obtained in~\cite{LaenenSterman}.
Near threshold, the dependence on rapidity is then entirely contained in 
the parton distribution functions: as can be seen from 
Eq.~(\ref{cross2}), their moments are shifted by $M$-dependent
terms. As a further source of rapidity dependence, one can keep the
Cosine term, as discussed above. It is straightforward to keep this term
when taking the Mellin moments, by writing
\begin{equation}
\cos\left( M\ln\frac{1}{\sqrt{z}}\right)=\frac{1}{2}
\left( z^{iM/2} + z^{-iM/2}\right) \; .
\end{equation}
This will simply result in a sum of two terms of the form~(\ref{c1qlargen})
with their moments shifted by $\pm iM/2$. 

We finally recall that contributions from the quark-gluon channels are 
not equally enhanced near threshold, but are all down by $1/N$. They
are therefore not subject to resummation, but will be included in our numerical 
studies at NLO level, using Eqs.~(\ref{app2})-(\ref{app5}).

\section{resummed cross section \label{sec4}}
In Mellin-moment space, threshold resummation for the
Drell-Yan process results in the exponentiation of the soft-gluon
corrections. To NLL the resummed formula
is given in the $\overline {{\mathrm{MS}}}$ scheme by~\cite{dyresum,dyresum2}
\begin{eqnarray}
&&\hspace*{-1.4cm}
\int_0^1 dz \,z^{N-1} \int_{-\ln\frac{1}{\sqrt{z}}}^
{\ln\frac{1}{\sqrt{z}}} d {\hat \eta} \, 
{\mathrm{e}}^{i M {\hat \eta}}\,x_1 x_2 D_{q\bar{q}}^{\mathrm{res}}
\left(x_1,x_2,\xaa,\xbb,\alpha_s(\mu^2),
\frac{M_W^2}{\mu^2}\right) \nonumber \\
&=&\exp\left[ C_q \left(\as(\mu^2),\ln  
\frac{M_W^2}{\mu^2}\right) \right]
\exp\left\{ 2 \int_0^1 d\zeta\, \f{\zeta^{N-1}-1}{1-\zeta}
\int_{\mu^2}^{(1-\zeta)^2 M_W^2} \f{dk_T^2}{k_T^2} A_q(\as(k_T^2))\right\}  ,
\label{dyres}
\end{eqnarray}
where
\begin{equation} \label{andim}
A_q(\as)=\frac{\as}{\pi} A_q^{(1)} +
\left( \frac{\as}{\pi}\right)^2 A_q^{(2)} + \ldots \; ,
\end{equation}
with~\cite{KT}:
\begin{equation}
\label{A12coef}
A_q^{(1)}= C_F
\;,\;\;\;\; A_q^{(2)}=\frac{1}{2} \; C_F  \left[
C_A \left( \frac{67}{18} - \frac{\pi^2}{6} \right)
- \frac{5}{9} N_f \right] \; .
\end{equation}
Here $N_f$ is the number of flavors and $C_A=3$.
The coefficient $C_q$ collects mostly
hard virtual corrections. Its exponentiation was
shown in~\cite{LaenenEynck}:
\begin{eqnarray}
C_q \left(\as(\mu^2),\ln  
\frac{M_W^2}{\mu^2}\right) =\frac{\as}{\pi}\,C_F\,
\left( -4+\frac{2\pi^2}{3}  +\frac{3}{2}\,\ln \frac{M_W^2}{\mu^2} \right)
 + {\cal O}(\as^2) \; .
\end{eqnarray}

Eq.~(\ref{dyres}) as it stands is ill-defined because
of the divergence in the perturbative running coupling
$\alpha_s(k_T^2)$ at $k_T=\Lambda_{\rm QCD}$. The perturbative
expansion of the expression shows factorial divergence,
which in QCD corresponds to a power-like ambiguity of the series.
It turns out, however, that the factorial divergence appears only
at nonleading powers of momentum transfer. The large logarithms
we are resumming arise in the region~\cite{dyresum2}
$\zeta\leq 1-1/\bar{N}$ in the integrand in Eq.~(\ref{dyres}). One therefore
finds that to NLL they are contained in the simpler expression
\begin{equation} \label{dyres1}
2 \int_{M_W^2/\bar{N}^2}^{M_W^2} \f{dk_T^2}{k_T^2} A_q(\as(k_T^2))
\ln\frac{\bar{N}k_T}{M}\,+\,
2 \int_{M_W^2}^{\mu^2} \f{dk_T^2}{k_T^2} A_q(\as(k_T^2))
\ln\bar{N}
\end{equation}
for the second exponent in~(\ref{dyres}).
This form is used for ``minimal''
expansions~\cite{Catani:1996yz} of the resummed exponent.

In the exponents, the large logarithms in $N$
occur only as single logarithms, of the form
$\as^k \ln^{k+1}(N)$ for the leading terms. Subleading terms
are down by one or more powers of $\ln(N)$. Knowledge of the
coefficients $A_q^{(1,2)}$ in Eq.~(\ref{dyres})
is enough to resum the full towers of LL terms
$\as^k \ln^{k+1}(N)$, and NLL ones $\as^k \ln^k(N)$ in the exponent.
With the coefficient $C_q$ one then gains
control of three towers of logarithms in the cross section,
$\as^k \ln^{2k}(N)$, $\as^k \ln^{2k-1}(N)$, $\as^k \ln^{2k-2}(N)$.

To NLL accuracy, one finds from 
Eqs.~(\ref{dyres}),(\ref{dyres1})~\cite{Catani:1996yz,CMN}
\be
\label{lndeltams}
&&\hspace*{-2.5cm}
\int_0^1 dz \,z^{N-1} \int_{-\ln\frac{1}{\sqrt{z}}}^
{\ln\frac{1}{\sqrt{z}}} d {\hat \eta} \, 
{\mathrm{e}}^{i M {\hat \eta}}\,x_1 x_2 D_{q\bar{q}}^{\mathrm{res}}
\left(x_1,x_2,\xaa,\xbb,\alpha_s(\mu^2),
\frac{M_W^2}{\mu^2}\right)\nonumber \\[3mm]
&=& \exp\left\{C_q \left(\as(\mu^2),\frac{M_W^2}{\mu^2}\right)
+2\ln \bar{N} \;h^{(1)}(\lambda) +
2 h^{(2)}\left(\lambda,\frac{M_W^2}{\mu^2}\right) \right\} \; ,
\ee
where
\be  \label{lamdef}
\lambda=b_0 \as(\mu^2) \ln \bar{N} \; .
\ee
The functions $h^{(1,2)}$ are given by
\label{h1fun}
\be
h^{(1)}(\la) &=& \f{A_q^{(1)}}{2\pi b_0 \la}
\left[ 2 \la+(1-2 \la)\ln(1-2\la)\right] \;,\\
h^{(2)}\left(\la,\frac{M_W^2}{\mu^2}\right)
&=&-\f{A_q^{(2)}}{2\pi^2 b_0^2 } \left[ 2 \la+\ln(1-2\la)\right]
+ \f{A_q^{(1)} }{2\pi b_0^3}
\left[2 \la+\ln(1-2\la)+\f{1}{2} \ln^2(1-2\la)\right]\nn \\
\label{h2fun}
&+& \f{A_q^{(1)}}{2\pi b_0}\left[2 \la+\ln(1-2\la) \right]
\ln\frac{M_W^2}{\mu^2}-
\f{A_q^{(1)}\as(\mu^2)}{\pi} \,\ln \bar{N}\, \ln\frac{M_W^2}{\mu^2} \;,
\ee
where
\begin{eqnarray}
b_0 &=& \frac{1}{12\pi} \left( 11 C_A - 2 N_f \right) \; , \nn \\
b_1 &=&  \frac{1}{24 \pi^2}
\left( 17 C_A^2 - 5 C_A N_f - 3 C_F N_f \right) \;\; .
\label{bcoef}
\end{eqnarray}
The function $h^{(1)}$ contains all LL terms in the perturbative
series, while $h^{(2)}$ is of NLL only.
When expanded to ${\cal O}(\alpha_s)$, Eqs.~(\ref{lndeltams})-(\ref{h2fun})
reproduce the full expression~(\ref{c1qlargen}) for the NLO
correction at large $N$. We note that the resummed exponent depends on the
factorization scales in such a way
that it will compensate the evolution
of the parton distributions. This feature is represented by
the last term in~(\ref{h2fun}). One therefore expects a decrease in
scale dependence of the cross section from resummation.
The remaining $\mu$-dependence in the second-to-last term
in~(\ref{h2fun}) results from the running of the strong coupling constant.

As was shown in Refs.~\cite{cat,KSV}, it is possible to improve
the above formula slightly and to also correctly take into account certain
subleading terms in the resummation. To this end, we rewrite
Eqs.~(\ref{lndeltams})-(\ref{h2fun}) as
\be
&&\hspace*{-2.5cm}
\int_0^1 dz \,z^{N-1} \int_{-\ln\frac{1}{\sqrt{z}}}^
{\ln\frac{1}{\sqrt{z}}} d {\hat \eta} \, 
{\mathrm{e}}^{i M {\hat \eta}}\,x_1 x_2 D_{q\bar{q}}^{\mathrm{res}}
\left(x_1,x_2,\xaa,\xbb,\alpha_s(\mu^2),
\frac{M_W^2}{\mu^2}\right)\nonumber \\[3mm]
&=&\exp\left\{
\frac{1}{\pi b_0}
\left[ 2 \la+\ln(1-2\la)\right]\,\left( \frac{A_q^{(1)}}{b_0 \alpha_s(\mu^2)}-
\frac{A_q^{(2)}}{\pi b_0}+\frac{A_q^{(1)}b_1}{b_0^2}+
A_q^{(1)}\,\ln \frac{M_W^2}{\mu^2}
\right) \right.\nonumber \\
&+&
\frac{\as(\mu^2)}{\pi}\,C_F\,
\left( -4+\frac{2\pi^2}{3}  \right)+
\frac{A_q^{(1)}b_1}{2\pi b_0^3} \,\ln^2(1-2\la)+
B_q^{(1)}\,\frac{\ln(1-2 \la)}{\pi b_0}
\nonumber \\
&+&\left. \left[-2 A_q^{(1)}\ln \bar{N}-B_q^{(1)}\right]\,
\left( \frac{\alpha_s(\mu^2)}{\pi} \ln \frac{M_W^2}{\mu^2} + 
\frac{\ln(1-2 \la)}{\pi b_0}
\right)\right\} \; , 
\label{lndeltams1}
\ee
where $B_q^{(1)}=-3 C_F/2$. The last term in Eq.~(\ref{lndeltams1})
is the LL expansion of the term
\be\label{evol}
\int_{\mu^2}^{M_W^2/\bar{N}^2} \f{dk_T^2}{k_T^2} \,\frac{\as(k_T^2)}{\pi} \,
\left[-2 A_q^{(1)}\ln \bar{N}-B_q^{(1)}\right]\; .
\ee
Since the factor $\left[-2 A_q^{(1)}\ln \bar{N}-B_q^{(1)}\right]$
is the large-$N$
limit of the moments of the LO splitting function $\Delta P^N_{qq}$, the 
term in Eq.~(\ref{evol}) may be viewed as the flavor non-singlet part of the 
evolution of the quark and anti-quark distributions between scales $\mu$ and
$M_W/\bar{N}$. This suggests to modify the resummation by
replacing~\cite{cat,KSV}
\be\label{evol1}
\left[-2 A_q^{(1)}\ln \bar{N}-B_q^{(1)}\right]\;\;\longrightarrow \;\;
C_F \left[ \frac{3}{2}-2 S_1(N)+{1\over N (N+1)}\right]
\ee
in Eq.~(\ref{lndeltams1}), the term on the right-hand-side
being the moments of the full LO non-singlet splitting 
function. With the help of this term, not only the leading 
large-$N$ pieces of the NLO $q\bar{q}'$ cross section are 
correctly reproduced, but also the contributions $\sim \ln(\bar{N})/N$.
We note that there are also related $\ln(\bar{N})/N$ pieces in the
NLO cross section for the $qg$ channel. These could be taken into 
account by extending~(\ref{evol1}) to the singlet evolution case, 
which we however refrain from in the present paper.

\section{Inverse transforms and matching \label{sec5}}

The final step in arriving at the resummed rapidity-dependent
cross section is to take the Mellin and Fourier inverse transforms of 
$\Delta\tilde{\sigma} (N,M)$ back to the variables $\tau$ and $\eta$:
\begin{eqnarray} \label{inverse} 
{d\Delta \sigma^{\mathrm{res}} \over d \eta} 
&=&\frac{1}{2 \pi} \int_{-\infty}^{\infty} dM \, {\mathrm{e}}^{-i M \eta} \, 
\frac{1}{2 \pi i} \int_{C-i \infty}^{C+i \infty} dN \, \tau^{-N} \,
\Delta \tilde{\sigma}^{\mathrm{res}}(N,M) \; . 
\end{eqnarray}
Care has to be taken when choosing the contour in complex $N$ space
because of the cut singularities in the resummed exponent. 
Adopting the ``minimal prescription'' of~\cite{Catani:1996yz} for the 
exponents, we 
choose the constant $C$ in~(\ref{inverse}) so that all singularities 
in the integrand are to the left of the integration contour, except for 
the Landau singularity at $N=\exp (1/2b_0 \alpha_s (\mu^2))$, 
which lies to the far right.
The contour is then deformed~\cite{Catani:1996yz} into the half-plane
with negative real part, which improves convergence while 
retaining the perturbative expansion. In this deformation, we need to 
avoid the moment-space singularities of the parton densities, which are 
displaced parallel to the imaginary axis by $\pm M/2$, as seen from 
Eq.~(\ref{cross2}). Thus, the intersection at $C$ of the contour with
the real axis has to lie far enough to the right that the contour 
does not pass through or below the singularities of the parton densities. 
The technique we use to achieve this is described in detail in
Ref.~\cite{sv}. 

In order to keep the full information contained in the NLO calculation,
we perform a ``matching'' of the NLL resummed cross section to the 
NLO one. This is achieved by subtracting from the resummed expression 
in Eq.~(\ref{inverse}) its ${\cal O}(\alpha_s)$ expansion,
\begin{equation} \label{inverse1} 
\frac{1}{2 \pi} \int_{-\infty}^{\infty} dM \, {\mathrm{e}}^{-i M \eta} \, 
\frac{1}{2 \pi i} \int_{C-i \infty}^{C+i \infty} dN \, \tau^{-N} \,
\left[ \Delta \tilde{\sigma}^{\mathrm{res}}(N,M) - 
\Delta \tilde{\sigma}^{\mathrm{res}}(N,M)\Big|_{{\cal O}(\alpha_s)}\right]\; ,
\end{equation}
and then adding the full NLO cross section, calculated using
Eqs.~(\ref{app1})-(\ref{app5}), which also includes the $qg$ channels.

\section{Phenomenological results \label{sec6}}

We are now ready to present some numerical results for our 
resummed rapidity-dependent cross sections and spin asymmetries
for $W^{\pm}$ production at RHIC. We are mainly aiming 
at investigating the quantitative effects of threshold resummation.
In how far the spin asymmetry for this process can provide
information on the polarized quark and anti-quark distributions
has amply been discussed in the 
literature~\cite{BOURRELY93,NadYuan2,grh,Chen:2005js}
and is not the focus of this work. We will therefore choose just
one set of polarized parton distributions of the proton, namely the NLO
``GRSV standard'' set of~\cite{grsv}. For the unpolarized parton 
distributions we take the NLO ones of~\cite{grv} throughout. Unless stated
otherwise, we choose the factorization and renormalization
scales as $\mu=M_W$.

\begin{figure}[t]
\centering
\includegraphics[width=8cm,height=8cm,clip]{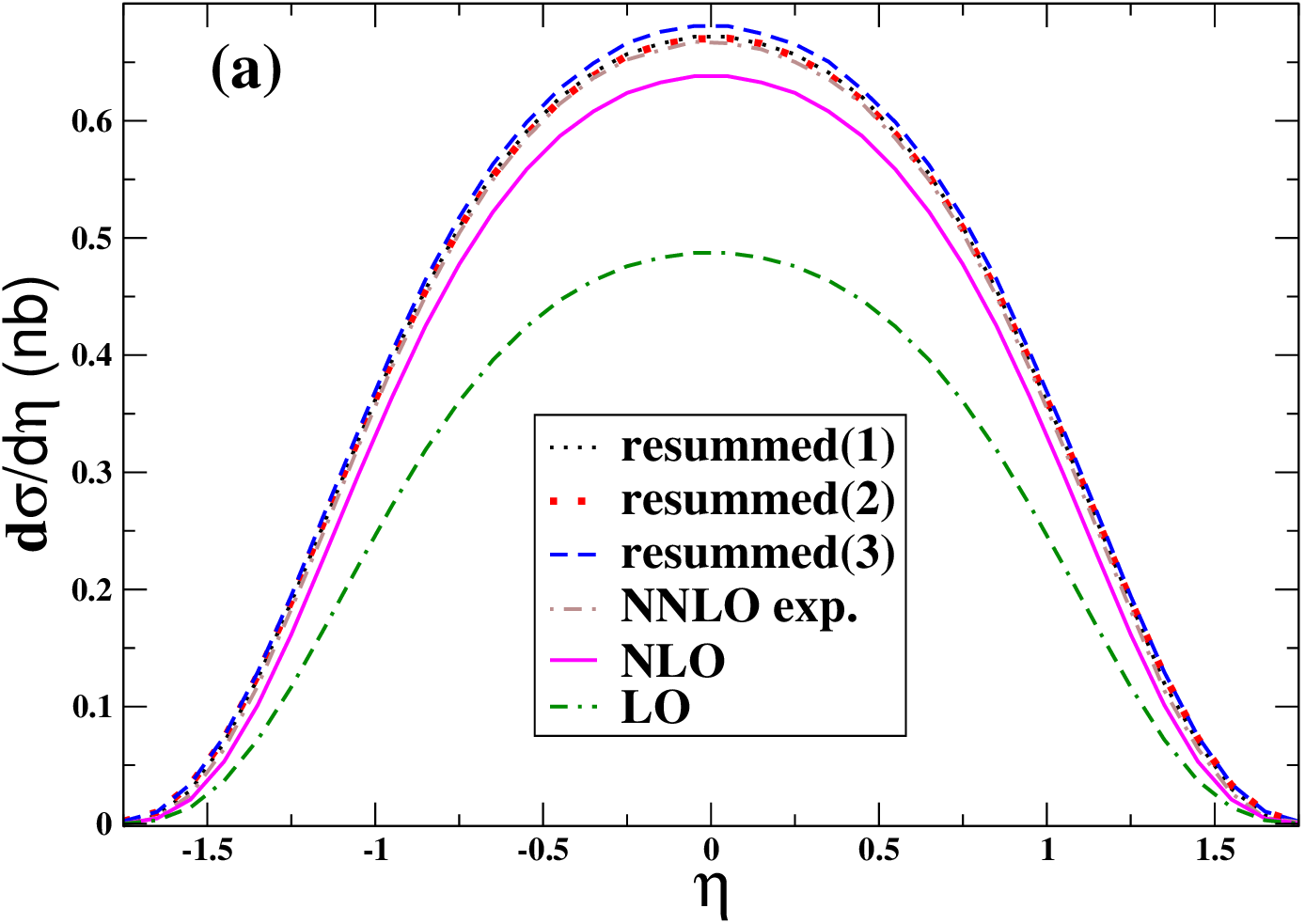}%
\hspace{0.2cm}
\includegraphics[width=8cm,height=8cm,clip]{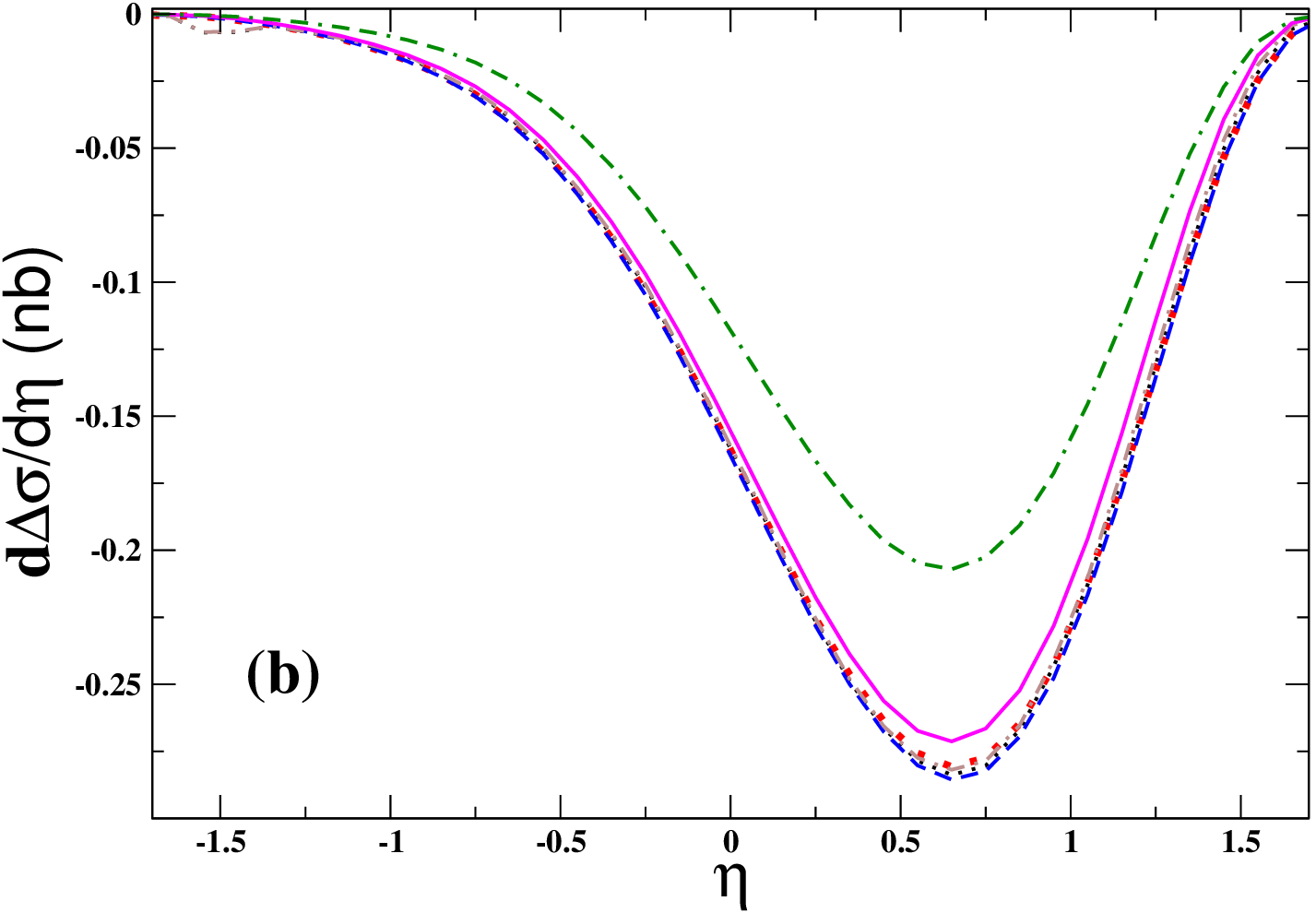}
\centering
\begin{minipage}[c]{0.5\textwidth}
\centering
\psfrag{B}{(c)}
\includegraphics[width=8cm,height=6cm,clip]{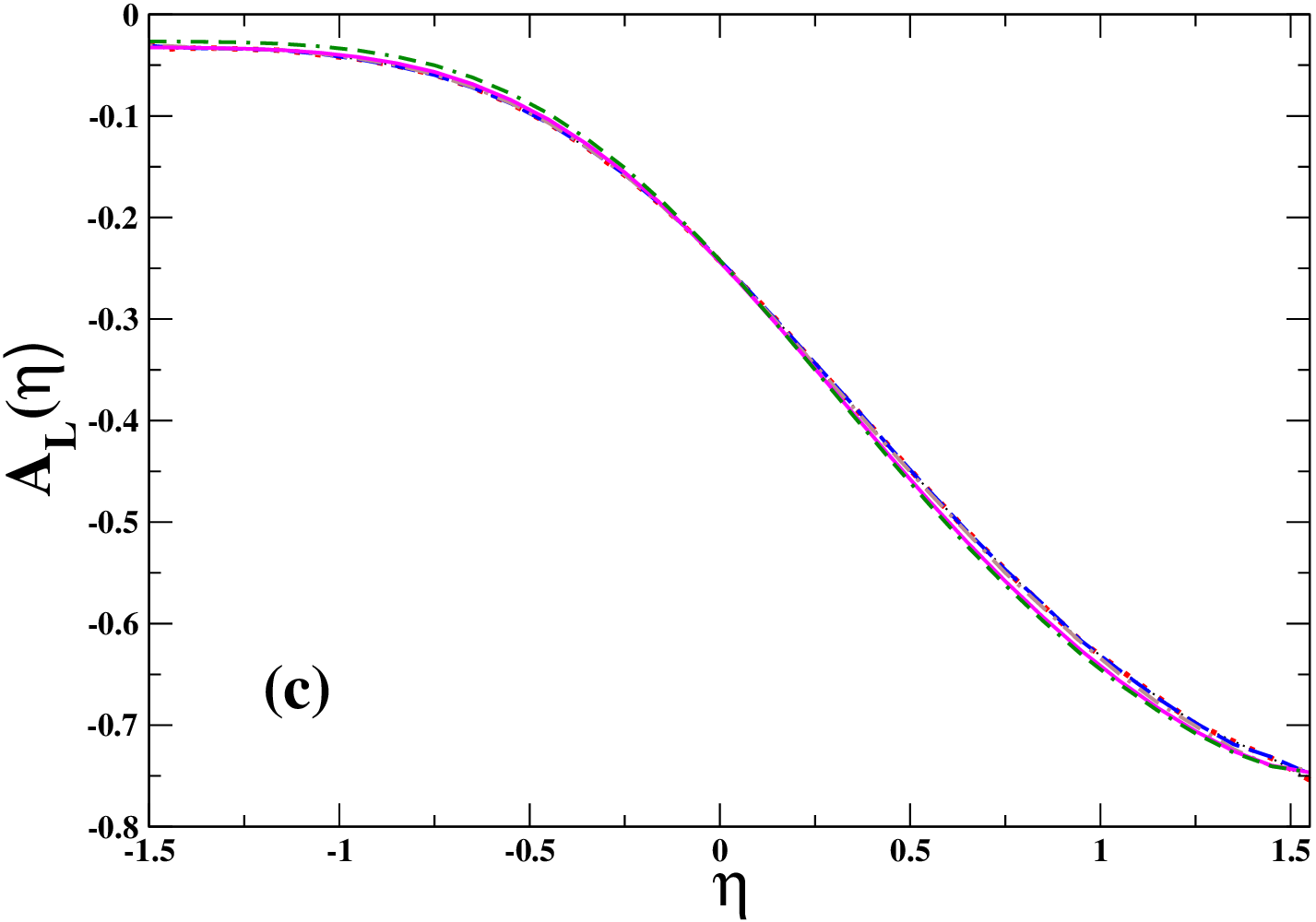}
\end{minipage}%
\caption{\label{fig1} (a) Differential 
cross section for $W^+$-production in
unpolarized $pp$ collision at RHIC for ${\sqrt S}=500 $~GeV as a function
of the rapidity of the produced boson. (b) Same, but
when one of the proton beams is longitudinally polarized. (c) The 
corresponding single-spin asymmetry ${\mathrm A}_L$. For all three plots, 
``resummed(1)'' denotes the threshold-resummed cross section without 
including subleading $\ln(\bar{N})/N$ terms and without the 
$\cos\left( M\ln\frac{1}{\sqrt{z}}\right)$ term, while for 
``resummed(2)'' the Cosine term is included. ``Resummed(3)'' is with
the $\ln(\bar{N})/N$ terms. The curve labelled ``NNLO exp.'' 
gives the NNLO expansion of the ``resummed(1)'' result. 
We also display the LO and NLO cross sections.}   
\end{figure}

Figures~\ref{fig1} (a) and (b) show the unpolarized and
single-spin polarized cross sections for $W^+$-boson production in 
$\sqrt{S}=500$~GeV collisions at RHIC, as functions of the  
rapidity of the boson. We show the results at various levels of 
perturbation theory. The lower line in Fig.~\ref{fig1} (a)
displays the unpolarized cross section at LO. The solid line above
is the NLO result. The lines referred to as ``resummed (1)-(3)'' 
are for the matched NLL resummed cross section. As one can see, they all lie a 
few per cent higher than the NLO cross section. 
For the two dotted lines, denoted as ``resummed (1)'' and ``resummed (2)'',
the subleading pieces $\propto \ln(\bar{N})/N$ have been neglected. For
``resummed (2)'' the $\cos\left( M\ln\frac{1}{\sqrt{z}}\right)$ term 
discussed after Eq.~(\ref{costerms}) is kept in the Mellin moment integrand, whereas
for ``resummed (1)'' it is ignored.
This evidently leads to a negligible difference.
The curve for ``resummed (3)'' then shows the effect of also including 
the subleading terms $\propto \ln(\bar{N})/N$. This leads to a 
further small increase of the predicted cross section. 
Finally, the remaining line, ``NNLO exp.'', represents
the two-loop (NNLO) expansion of the ``resummed (1)'' 
result. It shows that the NNLO terms generated by resummation are still 
significant, but that orders beyond NNLO have a negligible effect.
We note that we have checked that the NLO expansion of the resummed
cross section reproduces the exact NLO cross section very precisely.
This provides confidence that the logarithmic terms that are subject
to resummation indeed dominate the cross section, so that it is sensible
to resum them. We use the same line coding for the polarized 
cross section in Fig.~\ref{fig1} (b).

\begin{figure}[t]
\centering
\includegraphics[width=8cm,height=8cm,clip]{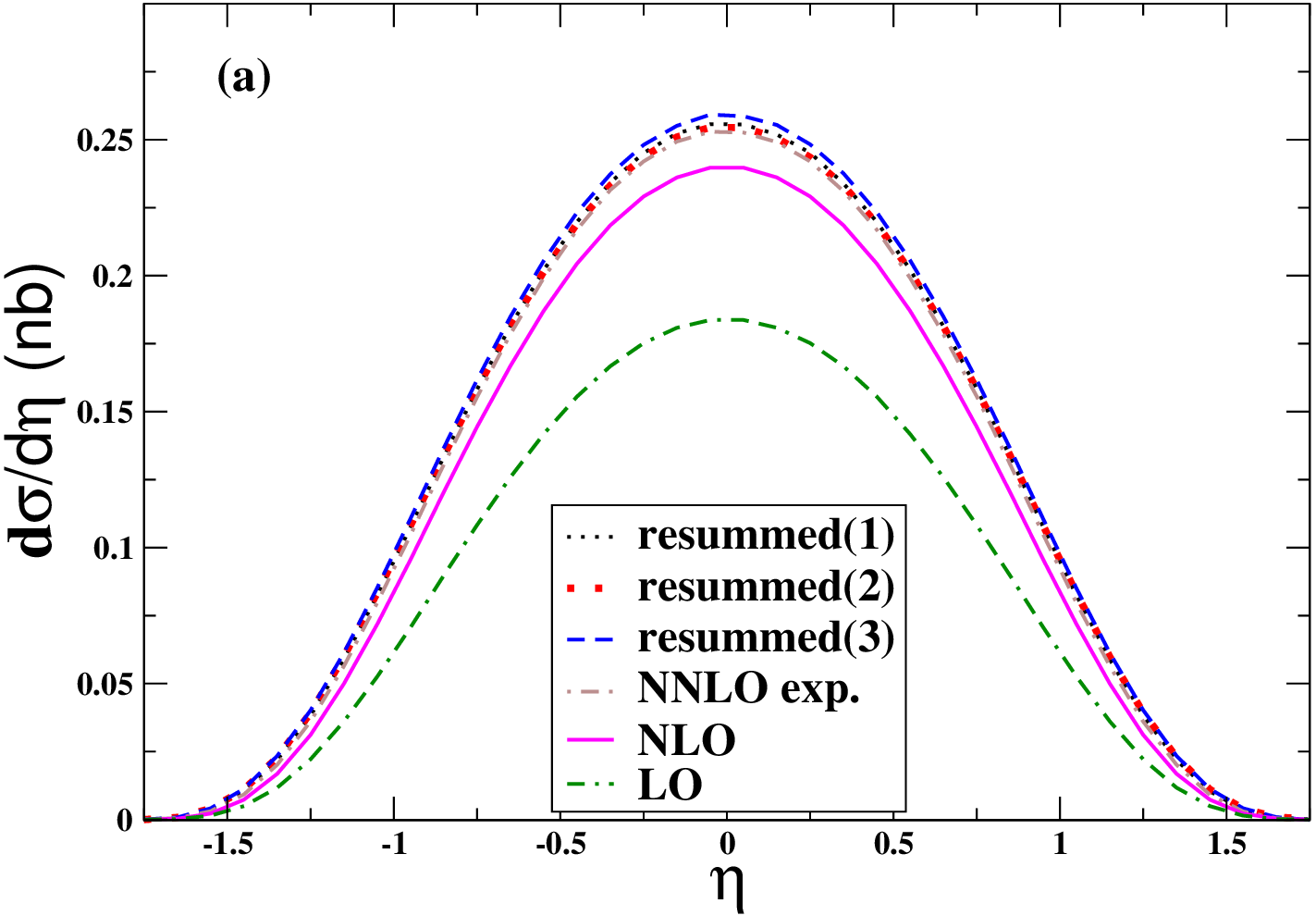}%
\hspace{0.2cm}
\includegraphics[width=8cm,height=8cm,clip]{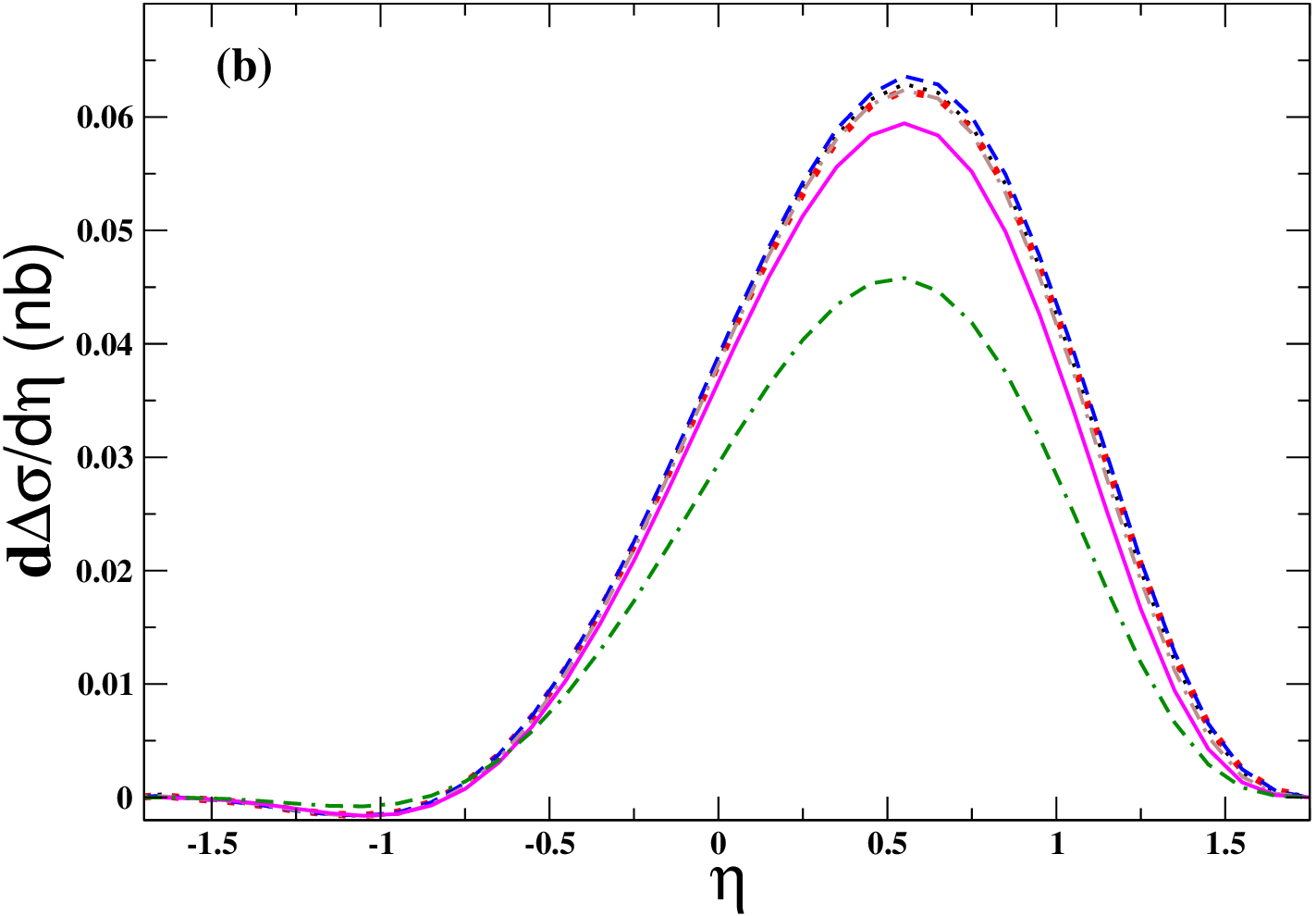}
\centering
\begin{minipage}[c]{0.5\textwidth}
\centering
\includegraphics[width=8cm,height=6cm,clip]{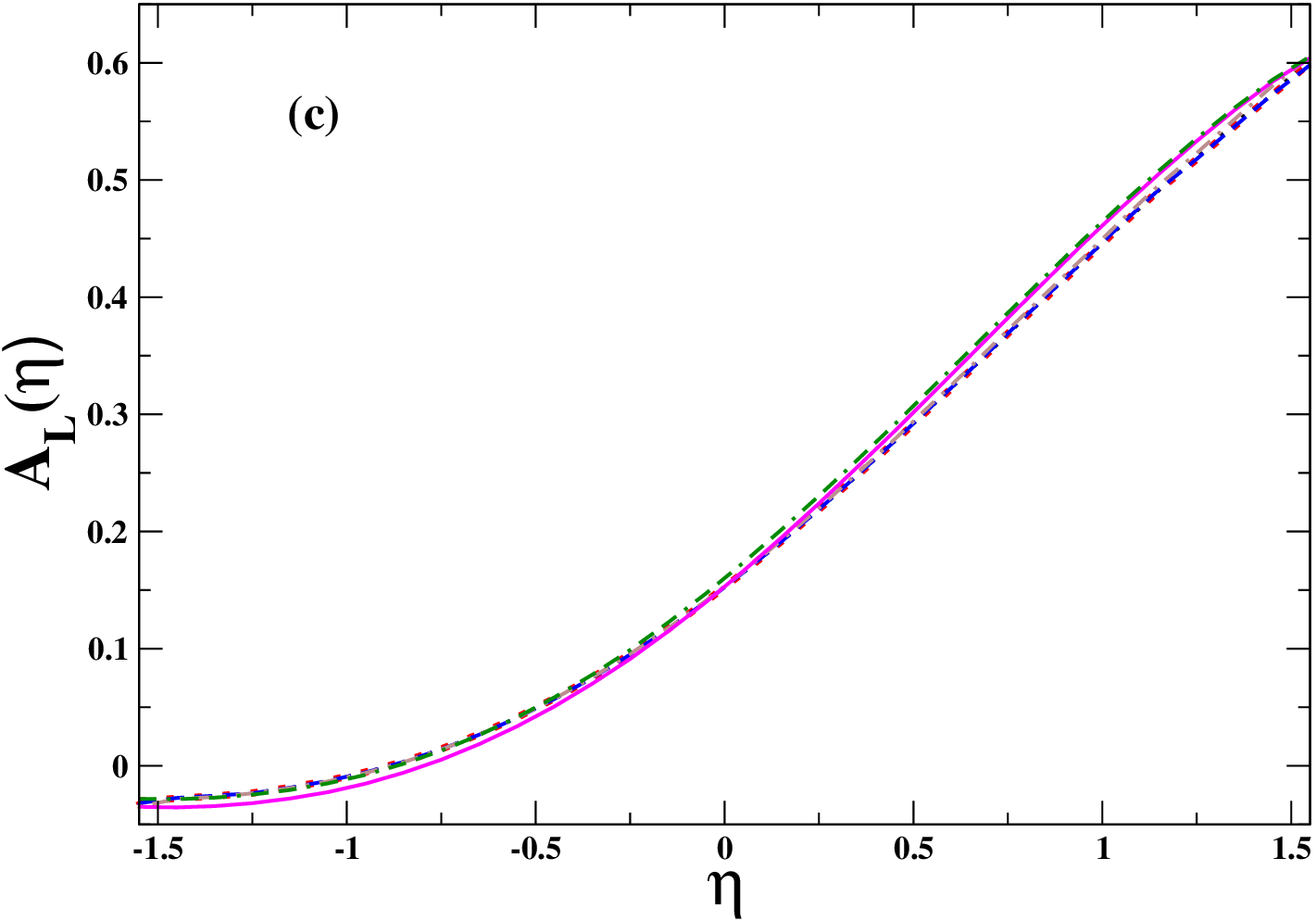}
\end{minipage}%
\caption{\label{fig2} Same as Fig.~\ref{fig1}, but for 
$W^-$ production.}
\end{figure}

Overall, we find rather moderate resummation effects at $\sqrt{S}=500$~GeV. 
In particular,
resummation does not affect the rapidity dependence of the cross
section much, in the sense that the shapes of the curves from NLO
to the NLL resummed case are very similar. This was to be
anticipated from our analytical results in Sec.~\ref{sec3}, where
we found that the resummation of the rapidity-dependent cross
section closely follows that of the rapidity-integrated one 
(see also~\cite{LaenenSterman}). Figure~\ref{fig1} (c) shows 
the corresponding single-spin asymmetry, obtained
by dividing the results in (a) and (b). The higher-order
effects cancel almost entirely, and resummation becomes
unimportant.

In Fig.~\ref{fig2} we display the corresponding results 
for $W^-$ production. Very similar quantitative features are
found, except of course that the cross sections are
smaller and, in the polarized case, of opposite sign
because of the different combinations of parton distributions involved.  

\begin{figure}[t]
\centering
\includegraphics[width=8cm,height=8cm,clip]{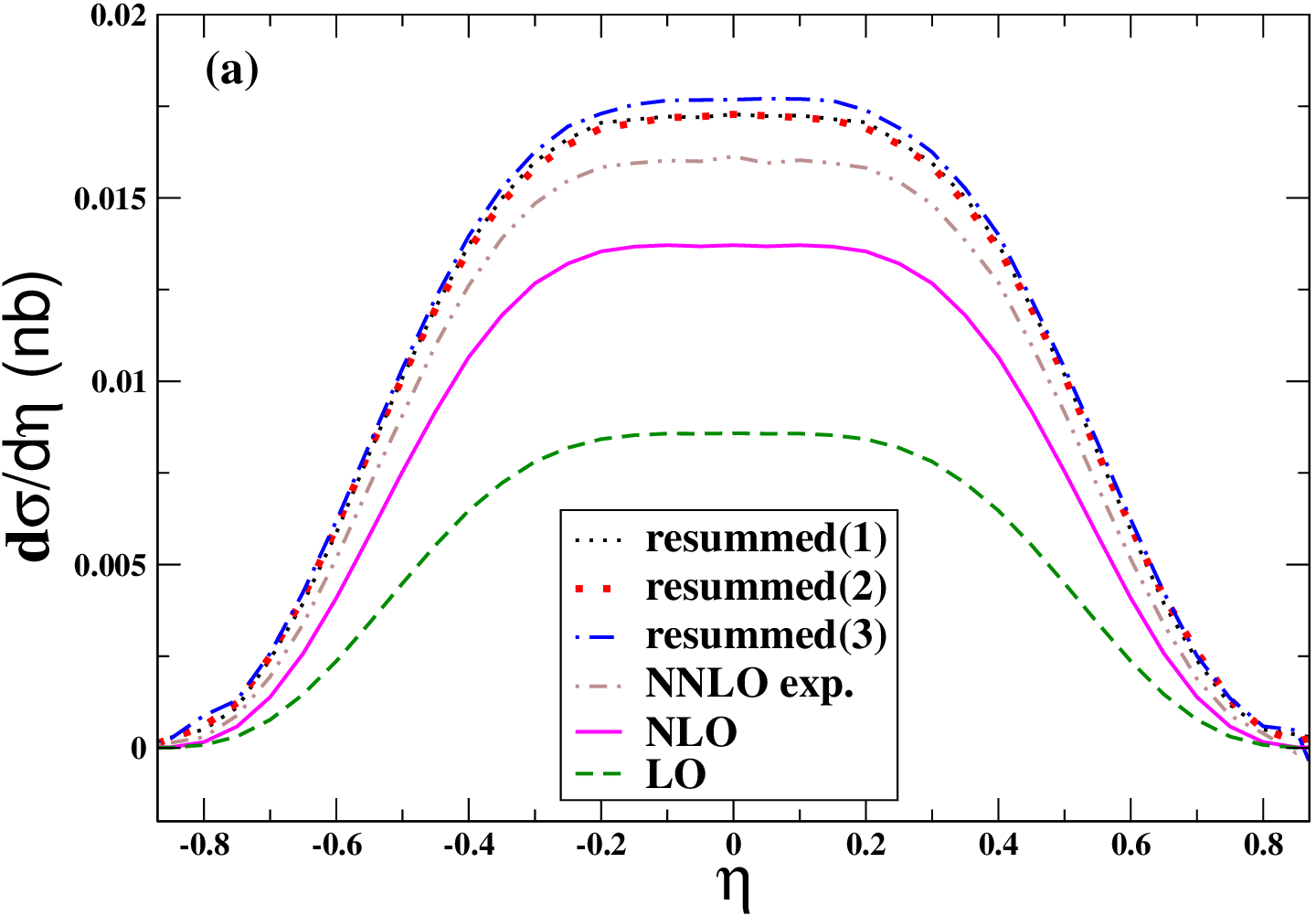}%
\hspace{0.2cm}
\includegraphics[width=8cm,height=8cm,clip]{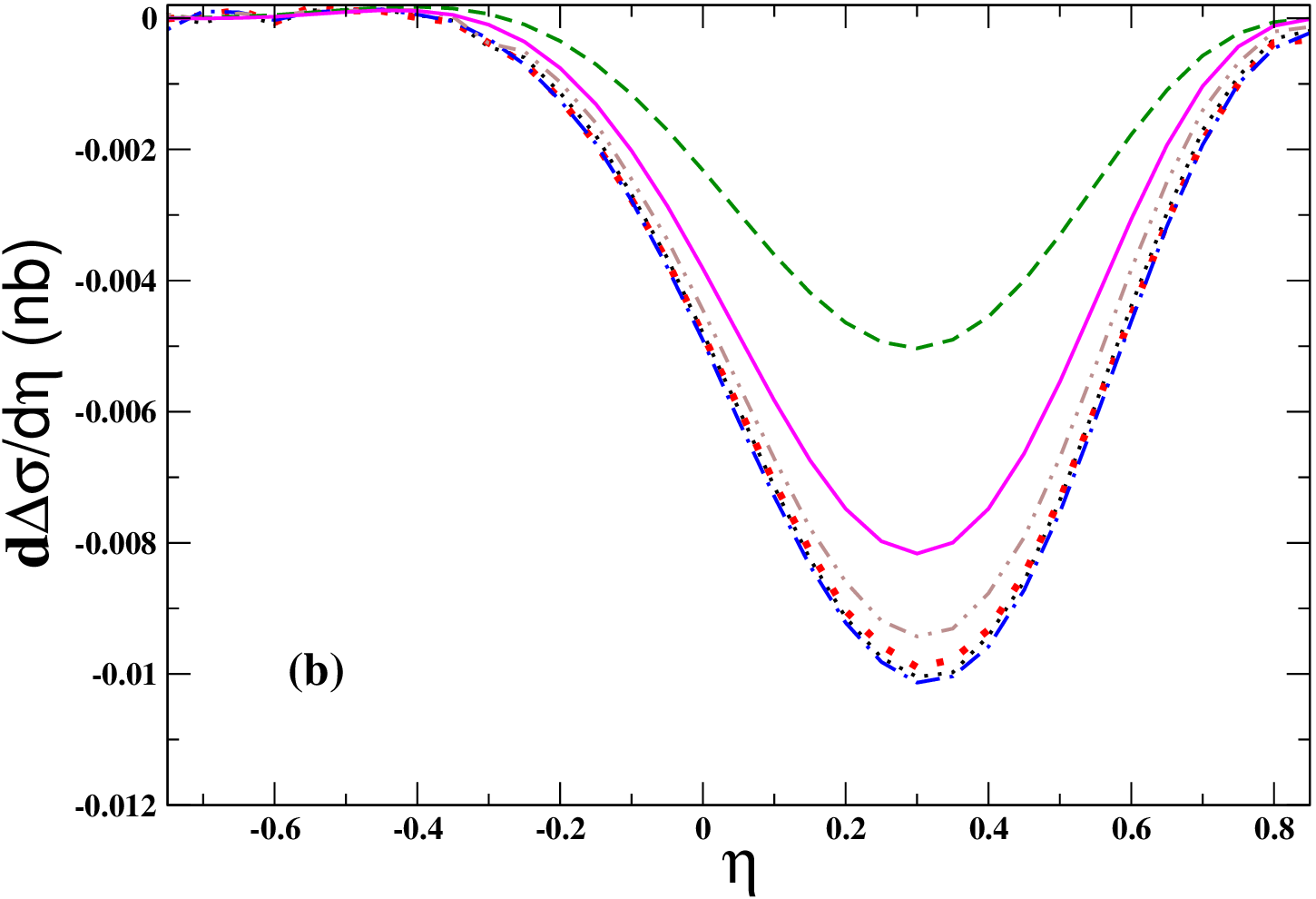}
\centering
\begin{minipage}[c]{0.5\textwidth}
\centering
\includegraphics[width=8cm,height=6cm,clip]{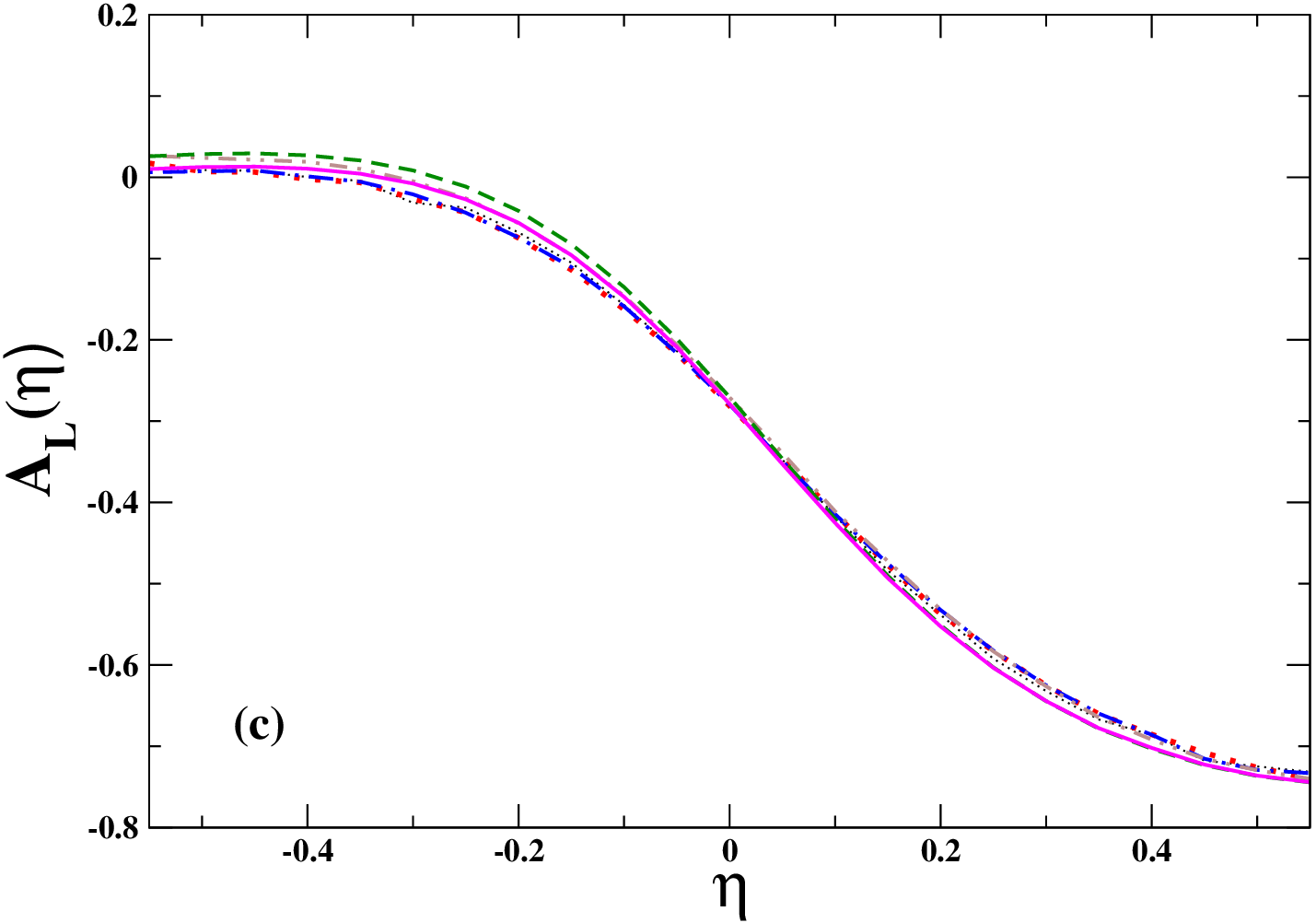}
\end{minipage}%
\caption{\label{fig3} Same as Fig.~\ref{fig1}, but for $\sqrt{S}=200$~GeV.}   
\end{figure}

\begin{figure}[t]
\centering
\includegraphics[width=8cm,height=8cm,clip]{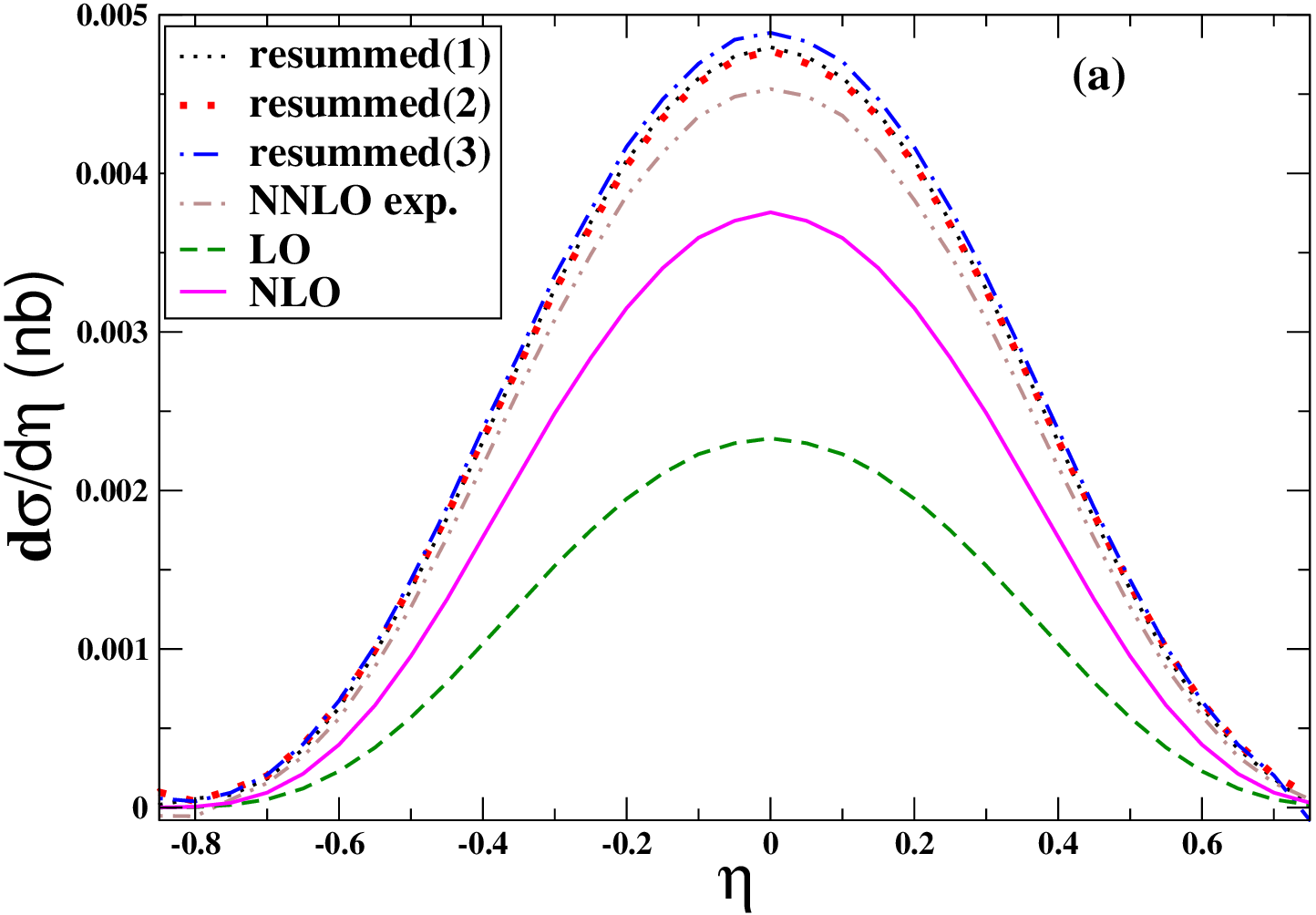}%
\hspace{0.2cm}
\includegraphics[width=8cm,height=8cm,clip]{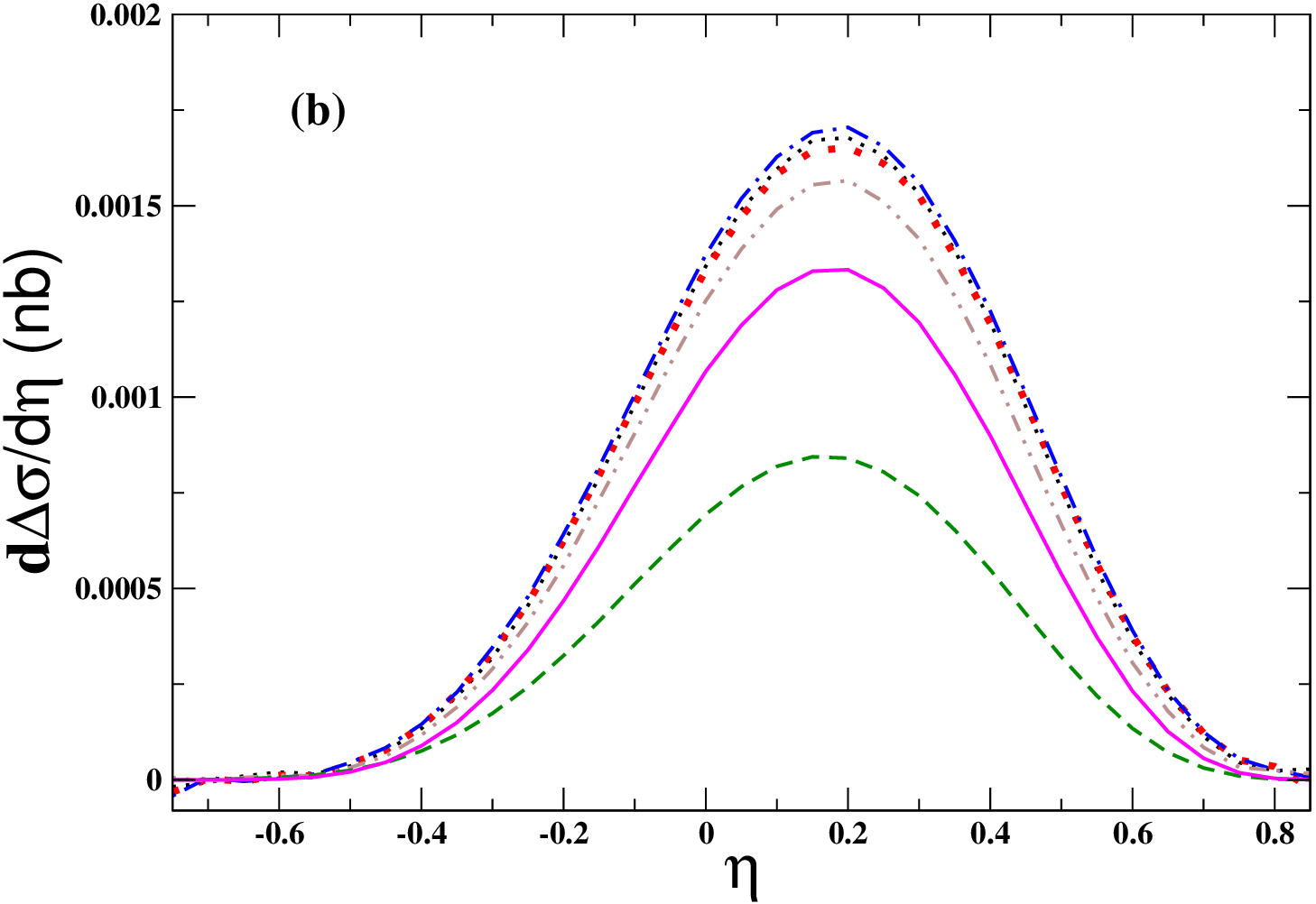}
\centering
\begin{minipage}[c]{0.5\textwidth}
\centering
\includegraphics[width=8cm,height=6cm,clip]{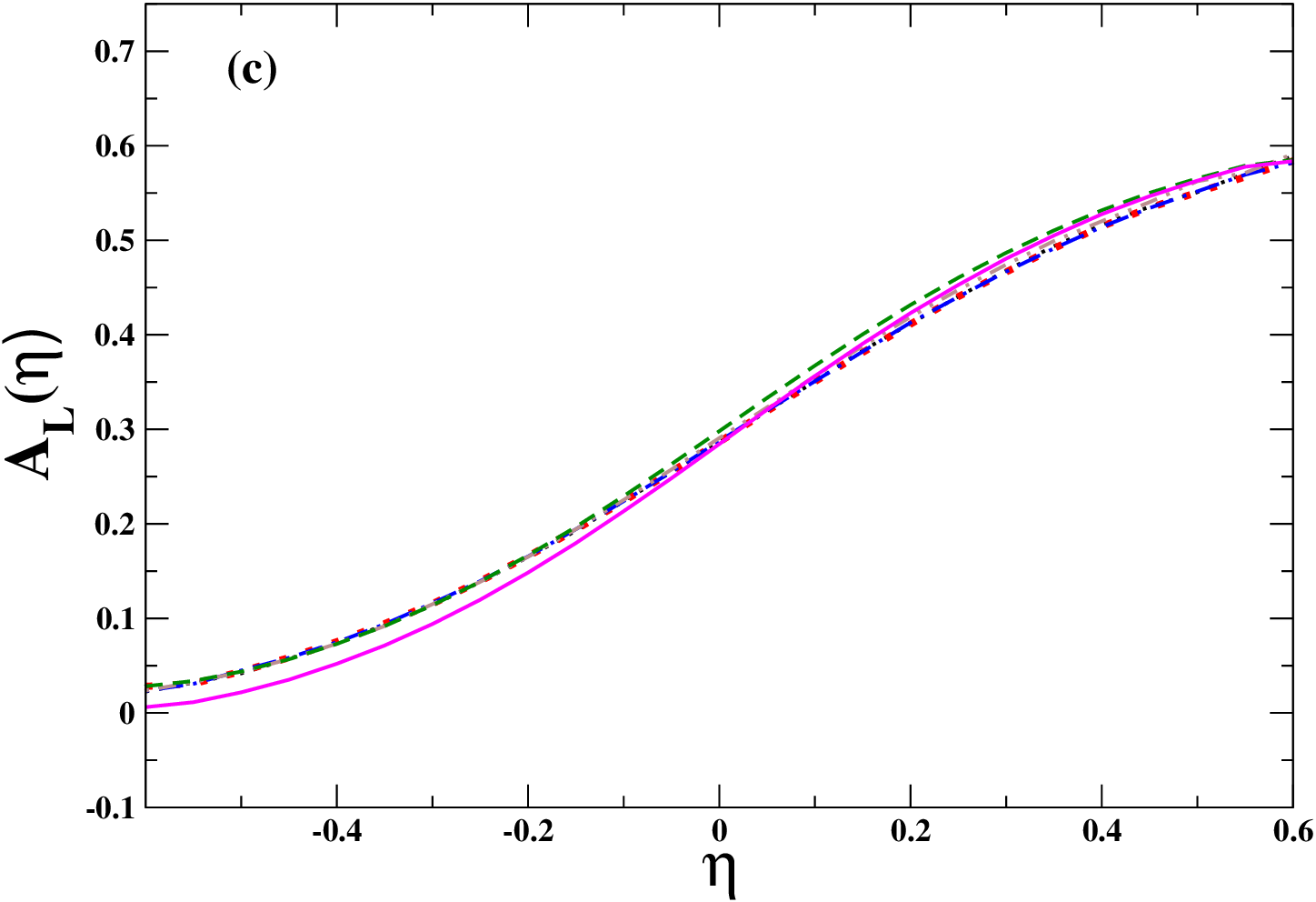}
\end{minipage}%
\caption{\label{fig4} Same as Fig.~\ref{fig2}, but for $\sqrt{S}=200$~GeV.} 
\end{figure}

Threshold resummation should become more important when $\tau$
increases. We thus expect larger perturbative corrections 
at RHIC's lower center-of-mass energy
of $\sqrt{S}=200$~GeV. In Figs.~\ref{fig3} and~\ref{fig4} we 
repeat our calculations in Figs.~\ref{fig1} and~\ref{fig2} at this
energy. Indeed, the resummation effects are much more significant, amounting
to about 25\% at mid rapidity. Also, since we are closer to threshold now,
subleading $\ln(\bar{N})/N$ terms are less important than before. 
Of course, the cross sections are much smaller at $\sqrt{S}=200$~GeV
than at 500~GeV. In fact, luminosity will need to be at least
around the design luminosity of $320$/pb in order to have 
sufficient statistics for a good measurement. It was shown recently~\cite{NadYuan2}
that at this luminosity a statistical accuracy of 5\% (9\%)
for the unpolarized $W^+$ ($W^-$) cross sections should be 
achievable, and that the current theoretical uncertainty related 
to the parton distributions is much larger, about 25\%. It may
therefore be quite possible to constrain better even the unpolarized parton 
distributions at large $x$ from measurements of $W^{\pm}$
production at RHIC at $\sqrt{S}=200$~GeV. Clearly, it will be
crucial then to take into account the perturbative corrections
we show in Figs.~\ref{fig3} and~\ref{fig4}, since these are
of similar size as the current parton distribution uncertainties.

\begin{figure}[t]
\centering
\includegraphics[width=8cm,height=8cm,clip]{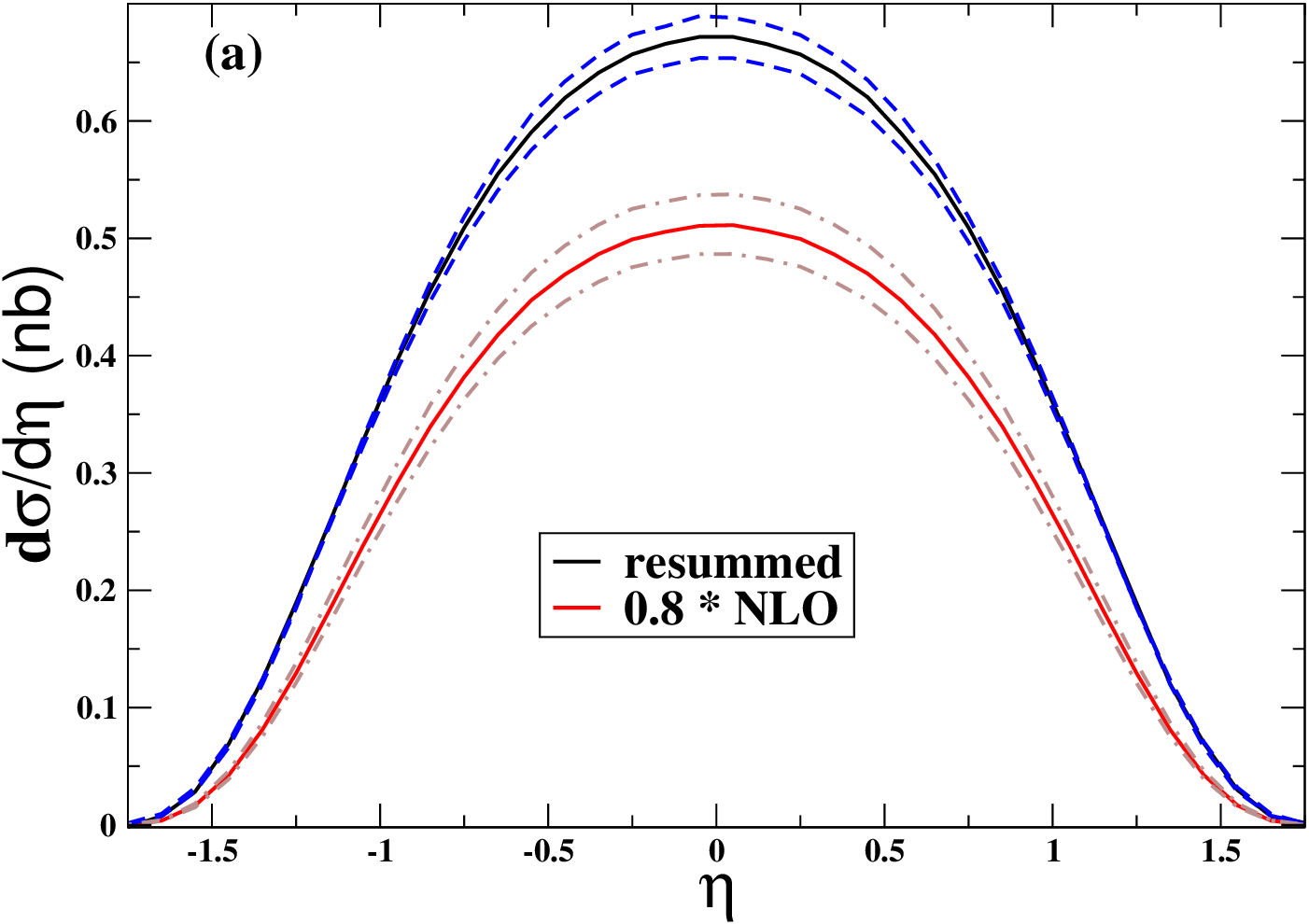}%
\hspace{0.2cm}
\includegraphics[width=8cm,height=8cm,clip]{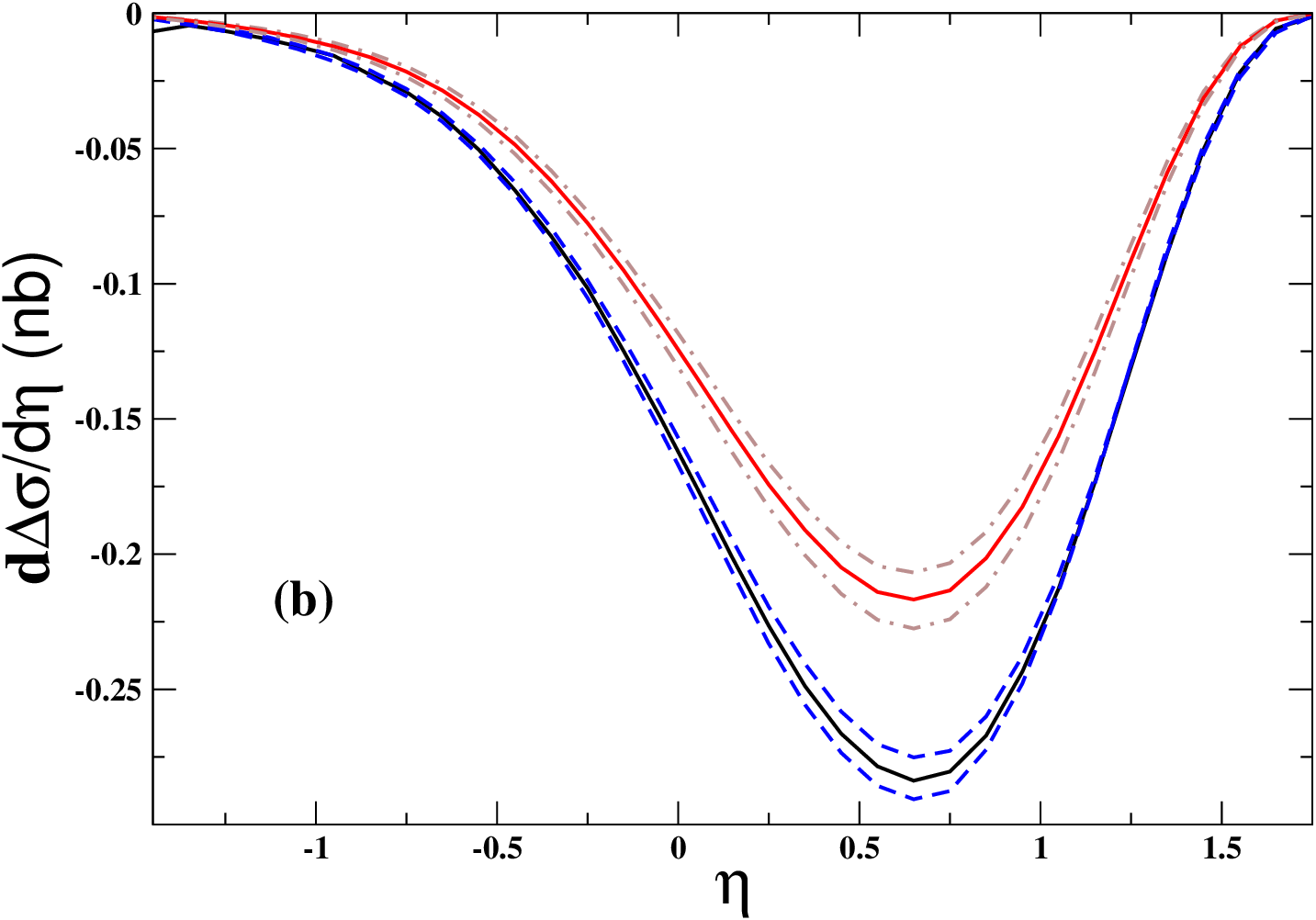}
\centering
\begin{minipage}[c]{0.5\textwidth}
\centering
\includegraphics[width=8cm,height=6cm,clip]{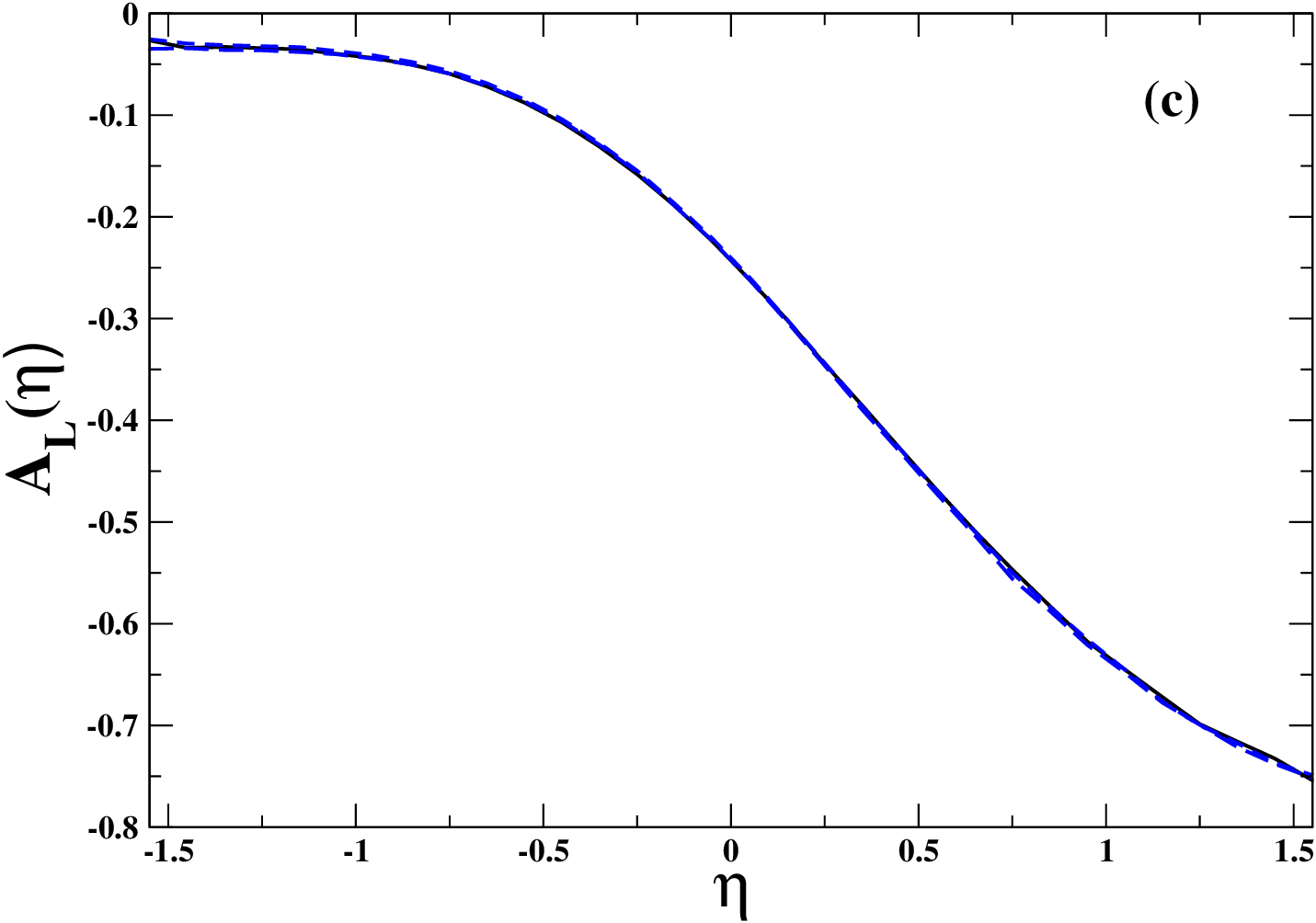}
\end{minipage}%
\caption{\label{fig5} (a) Scale dependence of the NLO and the 
``resummed(1)'' differential cross sections for $W^+$ production in
unpolarized $pp$ collisions at RHIC at ${\sqrt S}=500 $~GeV. 
We have varied the factorization/renormalization
scale $\mu$ between $M_W/2$ and $2 M_W$. Note that we have multiplied the NLO 
cross section by 0.8 for better visibility. The uppermost curves in each case
are for the scale $M_W/2$. (b) Same for the singly-polarized cross section. 
(c) Scale dependence of the single-spin asymmetry ${\mathrm A}_L$, calculated
from the resummed cross sections.}   
\end{figure}

So far, we have always chosen the factorization and renormalization
scales as $\mu=M_W$. As we discussed in Sec.~\ref{sec4}, threshold
resummation is expected to lead to a decrease of the scale dependence
of the cross sections. We finally examine this in Fig.~\ref{fig5},
where we vary the scales to $M_W/2$ and $2M_W$ in the NLO and the NLL 
``resummed(1)'' cross sections for $W^+$ production at $\sqrt{S}=500$~GeV.
Note that we have multiplied the NLO cross section by 0.8 for
better visibility. 
As expected, the scale dependence is improved after resummation. 
At NLL, it becomes very small, and it cancels almost entirely in the spin 
asymmetry.

\section{Conclusions and outlook \label{sec7}} 
We have performed a study of perturbative higher-order corrections
for $W^{\pm}$ production in singly-polarized $pp$ collisions at RHIC. 
This process will be used at RHIC to learn about the spin-dependent
$u$ and $d$ distributions of the proton, and about the corresponding 
anti-quark distributions. In this work we have dealt with the resummation of
potentially large ``threshold'' logarithms that arise when the incoming 
partons have just sufficient energy to produce the $W$ boson. We have performed
the resummation to next-to-leading logarithmic accuracy. We have considered
the resummation for the rapidity dependence of the cross sections,
for which we have used a method developed in~\cite{sv}. We find that
the resummation effect on the unpolarized and single-longitudinally polarized 
cross sections is rather moderate at RHIC's higher energy $\sqrt{S}=500$~GeV,
but more significant at $\sqrt{S}=200$~GeV where one is closer to threshold.
The shapes of the $W$ rapidity distributions are rather unaffected by resummation.
We believe that our study will be important in future extractions of 
the polarized, and even the unpolarized, $u,\bar{u},d,\bar{d}$ distributions
from forthcoming RHIC data. 

So far, we have only considered the $W^{\pm}$ as the observed final-state
particles. Ultimately, for the comparison to future data, it will be 
necessary to extend our study to a calculation of the cross section for a 
single charged lepton, since this is the signal accessible at RHIC. 
This will be devoted to a later study. We believe, however, that the results
presented here already give an excellent picture of the resummation
effects to be expected when the $W$ decay is fully taken into account. 

We finally note that the technique adopted in this work for treating 
the rapidity-dependence in the resummation for Drell-Yan type processes 
should also be useful for extending the recent study~\cite{yokoya}
of the Drell-Yan process in $p\bar{p}$ collisions at the 
GSI to the rapidity-dependent case. 

\section*{Acknowledgments}
This work originated from discussions at the ``8th Workshop on High Energy 
Physics Phenomenology (WHEPP-8)'', held at the Indian Institute of Technology,
Mumbai, January 2004. We are grateful to the organizers of the workshop and
to the participants of the working group on Quantum Chromodynamics for
useful discussions, in particular to R.\ Basu, E.\ Laenen, P.\ Mathews, 
A.\ Misra, and V.\ Ravindran. 
W.V.\ is grateful to RIKEN, Brookhaven National Laboratory
and the U.S.\ Department of Energy (contract number DE-AC02-98CH10886) for
providing the facilities essential for the completion of his work.

\appendix
\section{}
The NLO terms of the partonic coefficient functions in Eq.~(\ref{eq:lmas})
read~\cite{kubar,weber,tgdy,smrs}:
\begin{eqnarray}
D_{q\bar{q}}^{(1)}
\left(x_1,x_2,\xaa,\xbb,\frac{M_W^2}{\mu^2}\right)
& = & C_F G^A(x_1,x_2,x_1^0,x_2^0)\,\Bigg\{ \frac{2}{
\left[(\xa-\xaa)(\xb-\xbb)\right]_{+}} +
H^A (\xa,\xb,\xaa,\xbb)\nonumber \\
&& \delta (x_1-x_1^0) \,
\delta (x_2-x_2^0)\bigg[ \pi^2 - 8 + \ln^2
\frac{(1-x_1^0)(1-x_2^0)}{x_1^0 x_2^0}\bigg]  \nonumber \\
& & + \Bigg( \delta (\xa-\xaa) \bigg[ \frac{2 (x_2-x_2^0)}{x_2^2+x_2^0\, ^2}
+\frac{2}{x_2-x_2^0} \ln
\frac{2 x_2^0}{x_2+x_2^0} \nonumber \\
& & \hspace{0.6cm}
+ 2\left(\frac{\ln (\xb-\xbb)}{\xb-\xbb}\right)_{+}
+ \frac{2}{\left(\xb-\xbb\right)_{+}}\ln \frac
{1-\xaa}{\xaa\xbb}
\bigg] + (1 \leftrightarrow 2 ) \Bigg) \nonumber \\
&&
+ \ln \frac{M_W^2}{\mu^2} \Bigg\{  \delta (x_1-x_1^0)\,
\delta (x_2-x_2^0) \bigg[ 3 + 2 \ln\frac{(1-x_1^0)(1-x_2^0)}{x_1^0 x_2^0}
\bigg] \nonumber\\
&& \hspace{0.6cm}
+ \bigg( \delta (\xa-\xaa)\frac{2}{\left(\xb-\xbb\right)_{+}} + (1
\leftrightarrow 2 ) \bigg) \Bigg\} \Bigg\}\; ,   \label{app1} \\
D_{gq}^{(1)}
\left(x_1,x_2,\xaa,\xbb,\frac{M_W^2}{\mu^2}\right) & = &
T_F\Bigg\{\frac{\delta (\xb-\xbb) }{\xa^3} \Bigg[ (\xaa\,^2 +(\xa-\xaa)^2)
\ln\frac{2(\xa-\xaa)(1-\xbb)}{(\xa+\xaa)\xbb}
 \nonumber \\
& & \hspace{0.6cm}
+ 2\xaa(\xa-\xaa)\Bigg]
+\frac{G^C(\xa,\xb,\xaa,\xbb)}{(\xb-\xbb)_{+}} +
H^C(\xa,\xb,\xaa,\xbb)
\nonumber \\ & &
+  \ln \frac{M_W^2}{\mu^2} \Bigg\{ \frac{\delta (\xb-\xbb) }{\xa^3}
 (\xaa\,^2 +(\xa-\xaa)^2) \Bigg\}\Bigg\}\; , \label{app2}\\
D_{qg}^{(1)}
\left(x_1,x_2,\xaa,\xbb,\frac{M_W^2}{\mu^2}\right) & = &
D_{gq}^{(1)}
\left(x_2,x_1,\xbb,\xaa,\frac{M_W^2}{\mu^2}\right) \; , \label{app3} \\
\Delta D_{gq}^{(1)}
\left(x_1,x_2,\xaa,\xbb,\frac{M_W^2}{\mu^2}\right) & = &
T_F\Bigg\{\frac{\delta (\xb-\xbb) }{\xa^2} \Bigg[ (2\xaa -\xa)
\ln\frac{2(\xa-\xaa)(1-\xbb)}{(\xa+\xaa)\xbb}
 \nonumber \\
& & \hspace{0.6cm}
+ 2(\xa-\xaa)\Bigg]
+ \frac{\Delta G^C(\xa,\xb,\xaa,\xbb)}{(\xb-\xbb)_{+}} +
H^C(\xa,\xb,\xaa,\xbb)
\nonumber \\ & &
+  \ln \frac{M_W^2}{\mu^2} \Bigg\{ \frac{\delta (\xb-\xbb) }{\xa^2}
 (2\xaa -\xa) \Bigg\}\Bigg\}\; , \label{app4} \\
\Delta D_{qg}^{(1)}
\left(x_1,x_2,\xaa,\xbb,\frac{M_W^2}{\mu^2}\right) & = &
\Delta D_{gq}^{(1)}\left(x_2,x_1,\xbb,\xaa,\frac{M_W^2}{\mu^2}\right)\; ,
\label{app5}
\end{eqnarray}
where
\begin{eqnarray}
G^A(\xa,\xb,\xaa,\xbb) & = & \frac{(\xa\xb+\xaa\xbb)
(\xaa\,^2\xbb\,^2+\xa^2\xb^2)}
{\xa^2\xb^2(\xa+\xaa)(\xb+\xbb)}\; , \nonumber \\
H^A (\xa,\xb,\xaa,\xbb) & = &
-\frac{4\xa\xb\xaa\xbb(\xa+\xaa)(\xb+\xbb)}{(\xaa\xb+\xa\xbb)^2
\left( \xa^2\xb^2+\xaa\, ^2\xbb\, ^2 \right)}\; , \nonumber \\
G^C(\xa,\xb,\xaa,\xbb) & = & \frac{2\xbb(\xaa\,^2\xbb\,^2+(\xaa\xbb-\xa\xb)^2)
(\xaa\xbb+\xa\xb)}{\xa^3\xb^2(\xa\xbb+\xb\xaa)(\xb+\xbb)}\; ,
\nonumber \\
H^C (\xa,\xb,\xaa,\xbb) & = &\frac{2\xaa\xbb(\xaa\xbb+\xa\xb)(\xa\xaa\xb^2+
\xaa\xbb(\xa\xbb+2\xaa\xb))}{\xa^2\xb^2(\xa\xbb+\xb\xaa)^3}\; , \nonumber\\
\Delta G^C(\xa,\xb,\xaa,\xbb)
& = & \frac{2\xbb(2\xaa\xbb-\xa\xb)(\xaa\xbb+\xa\xb)}
{\xa^2\xb(\xa\xbb+\xb\xaa)(\xb+\xbb)}\; .
\label{app6}
\end{eqnarray}
Here the ``plus''-distributions are defined as
\begin{equation}
\int_{x_1^0}^1 dx_1 \frac{f(x_1)}{(x_1-x_1^0)_+}\equiv
\int_{x_1^0}^1 dx_1 \frac{f(x_1)-f(x_1^0)}{x_1-x_1^0} 
\end{equation}
(and likewise for $x_2$), and 
\begin{eqnarray}
&&\hspace*{-2cm}\int_{x_1^0}^1 dx_1 \int_{x_2^0}^1 dx_2 \frac{g(x_1,x_2)}{\left[
(x_1-x_1^0)(x_2-x_2^0)\right]_+}\nonumber \\ &\equiv&
\int_{x_1^0}^1 dx_1 \int_{x_2^0}^1 dx_2 \frac{g(x_1,x_2)-g(x_1^0,x_2)-
g(x_1,x_2^0)+g(x_1^0,x_2^0)}{(x_1-x_1^0)(x_2-x_2^0)} \; .
\end{eqnarray}


\end{document}